# Dual Function of Plasma-Activated Water in Lettuce: Growth Promoter and Nitrogen Source under Nitrogen-sufficient/-deprived Conditions

Adriana Mišúthová[1], Zuzana Lukačová[2], Ramin Mehrabifard[1], Zdenko Machala[1]

[1]Division of Environmental Physics, Faculty of Mathematics, Physics and Informatics, Comenius University, Bratislava, Slovakia

[2]Department of Plant Physiology, Faculty of Natural Sciences, Comenius University, Bratislava, Slovakia

E-mail: adriana.misuthova@fmph.uniba.sk

## Abstract

Cold atmospheric plasma generates reactive oxygen and nitrogen species (RONS) that dissolve in water, forming plasma-activated water (PAW) and inducing complex chemical changes and biological responses. PAW, enriched with nitrogen (N) species, emerges as a sustainable alternative to fertilisers by combining signalling (ROS) and N supplementation (RNS). This study aims to investigate the dual role of PAW in lettuce, *Lactuca sativa* (cv. Matilda), grown under either N-sufficient or N-deprived half-strength Hoagland solutions. Four treatments were evaluated: control, seed priming in PAW-activated Hoagland, PAW-activated Hoagland cultivation, and the combination of the latter two, under both N-content regimes.

Under N-sufficient conditions, seed priming in PAW increased the lettuce head fresh weight up to 40%, chlorophyll *a+b* by 25%, and the sensory scores (freshness, juiciness), compared to the control. Under N-deprived conditions, PAW partially compensated for N-deficiency, increased the plant growth up to 50%, photosynthetic pigments and quercetin (up to 20%), and antioxidant enzymes (e.g., APX and G-POX increased up to 2x). Principal component analysis confirmed N-content as the primary variance driver (PC1 69–71%), with PAW modulating redox homeostasis tissue-specificity.

Our results demonstrate the capacity of PAW to enhance plant growth, metabolism, and quality, even in N-deprived conditions, to address gaps in plasma agriculture by highlighting the PAW nutritional and signalling contributions. PAW thus represents a promising, eco-friendly strategy.

Key words: antioxidant enzymes, lettuce, nitrogen deficiency, plasma-activated water, RONS, secondary metabolites, sensory analysis

## Introduction

Plants are sessile organisms that depend on their environment. The soil solution is the primary source of essential nutrients, which are taken up primarily through the plant root system. One of the most important macronutrients for plant growth and development is nitrogen (N). A major source of this element in soil solution is nitrate, which is derived from inorganic minerals. Due to the high plant demand for this macronutrient, global agriculture is dependent on its supplementation in the form of inorganic fertilisers (Erisman et al., 2008). As the price of chemical fertilisers gradually increases, carbon emissions from their production significantly contribute to the climate change, alternative methods for delivering nutrients to the plant body are being investigated (Smith et al., 2008). Another downside of using synthetic fertilisers is the risk of low quality and lack of control, which often results in soil contamination with heavy metals (Mortvedt, 1995).

A promising alternative for producing nitrate from air, rather than using chemical fertilisers, are plasma-based techniques. Plasma, the fourth state of matter, is a source of electrons and positively charged ions, photons, and several atomic and molecular radicals (Randeniya and De Groot, 2015). In addition to natural lightning phenomena on Earth, it can be generated in the laboratory by providing high-energy excitation to gas molecules.

The so-called nonthermal (cold) plasma with elevated electron energy but low gas temperature is currently widely used in agriculturally focused research (Bilea et al., 2024; Kizer et al., 2025; Okruhlicová et al., 2025; Puač et al., 2018; Ruamrungsri et al., 2023). Highly valuable molecules for plant metabolism are produced within water after its activation by plasma (PAW), especially reactive oxygen and nitrogen species (RONS). These highly reactive radicals and molecules act as double-edged swords: at higher concentrations, they impose oxidative stress on cells, whereas at lower concentrations, they mitigate stress responses and are involved in plant signalling (Alam et al., 2025). Most importantly, RONS production in plasma-activated water can be controlled. In this way, PAW can help to activate developmental processes, hormonal responses(Ghulam et al., 2025) , and defence mechanisms in plant metabolism (Adhikari et al., 2019; Lukacova et al., 2021; Zambon et al., 2020), rather than acting destructively.

ROS are recognised as second messengers in cell signalling that play an essential role in a wide range of cellular processes. A considerable attention is given to them in the induction of tolerance mechanisms under stress (Faraz et al., 2023; Martin et al., 2022). Plant cells and tissues produce ROS as byproducts of their metabolism under physiological conditions; however, under stress, a delicate balance between their generation and scavenging can be destroyed, thus leading to oxidative stress or even oxidative damage, manifesting as chlorophyll bleaching, which means impairing photosynthetic efficiency with many negative consequences for plant fitness. Other consequences are increased lipid peroxidation, membrane injury, DNA damage or reduced protein content (Kumari et al., 2025). Plant status depends on the cell's ability to produce effective enzymatic and non-enzymatic antioxidants, thereby maintaining ROS homeostasis. Among them, peroxidases (POX), superoxide dismutases (SOD), catalases (CAT), glutathione reductase (GR) are the most important enzymatic antioxidants, and carotenoids and phenolic compounds belong to nonenzymatic antioxidants (Dvořák et al., 2021; Hunt et al., 2021)

Hydroponic cultivation represents one of the most widely used soilless growing systems, alongside other approaches such as nutrient film technique (NFT), aeroponics, and deep-water culture systems. These systems differ in nutrient delivery strategy and root zone aeration (Elakrouch et al., 2026). In this study, we focus on several factors determining the efficiency of reactive oxygen and nitrogen species (RONS), particularly $NO_3^-$, $NO_2^-$, and $H_2O_2$, in water after activation by cold atmospheric plasma, which generates plasma-activated water (PAW), and their effects on *Lactuca sativa* plants cultivated using a horizontal deep-water hydroponic system with continuous aeration of the nutrient solution. We investigate the stability of RONS in PAW and assess whether PAW can act as a partial or full nitrogen substitute without external nitrogen supplementation under nitrogen-deprived conditions. Secondly, we focus on the potential seed priming effect of $H_2O_2$ following PAW treatment of *L. sativa* seeds. We aim to determine whether exogenously derived $H_2O_2$ acts as a signalling molecule, leading to activation of the antioxidant defence system, or whether it induces a shift in redox homeostasis associated with a stress response. Finally, we evaluate whether PAW treatment influences the nutritional quality of lettuce, with particular emphasis on quercetin content, and assess its potential impact on sensory perception by consumers.

## Materials and Methods

### Preparation of PAW - Multi-needle transient spark discharge

For preparing plasma-activated water, we used a multi-needle transient spark (TS) discharge. Transient spark is a highly energetic non-thermal plasma source of RONS that can use water surface as one of the electrodes (Mehrabifard et al., 2024). This multineedle TS plasma setup has been extensively used in our research group (Mehrabifard et al., 2025) (Ghulam et al., 2025). The TS system operates in pin-to-water geometry with a common liquid-grounded electrode. The power needle electrodes are connected to a DC power supply, each separately through a 4.7 MΩ ballast resistor and 50 pF external capacitor. PAW was prepared by treating 5 L of tap water with 21 parallel TS discharges over 238 minutes. There are two water containers: 1-litre container under the electrodes and a 4-litre container at the bottom. The water is circulated through the system by a pump at a constant flow rate (~0.4 L $min^{-1}$). Water is transferred from one side to the upper container where it is treated by the plasma discharges and returns to the larger container. Each TS operated at the repetitive frequency of spark pulses ~1 kHz with electrodes 1 cm above the water surface. Figure 1 A and B show a schematic diagram and a photo of the multi-needle TS plasma source, and Figure 1 C and D show the voltage and current-voltage waveform of the discharge, respectively. The TS plasma power was measured using voltage, current, and frequency. The instantaneous power, P(t), at any time t is given by:

$$P(t) = V(t).I(t)$$

Where V(t) is the instantaneous voltage, and I(t) is the instantaneous current. To find the average power over one pulse period, we integrate P(t) over the period (T) and divide by T, i.e. multiply by the frequency (f):

$$P_{avg} = \frac{1}{T}\int_0^T P(t)dt = f\int_0^T V(t).I(t)\,dt$$

The typical average power of the Multipin TS is 52.5 W, i.e. 2.5 W per needle.

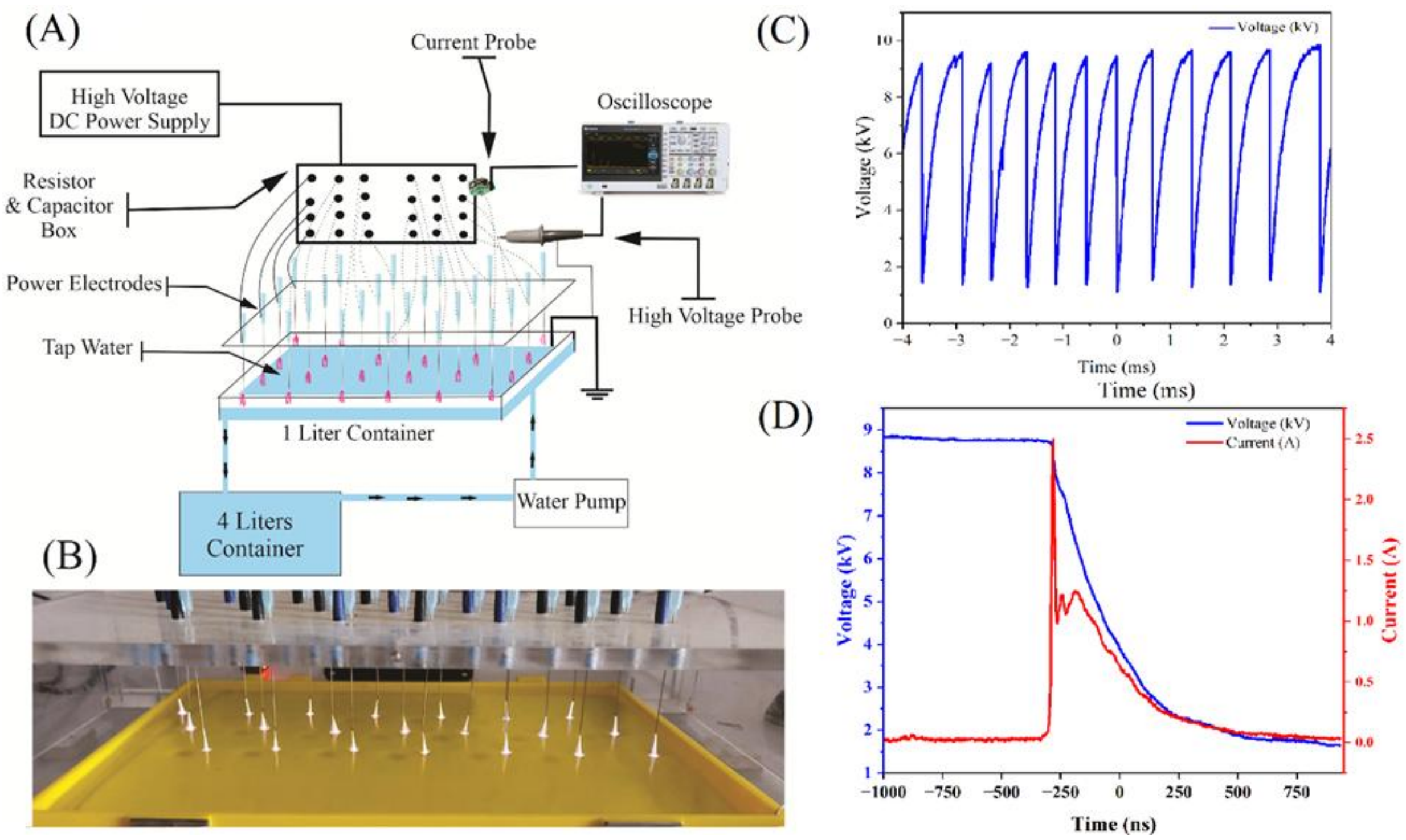

*Fig. 1 Multi-needle TS discharge, (A) schematic diagram, (B) real photograph, (C) voltage waveform on ms timescale, and (D) current-voltage waveform on ns timescale.*

### ***Chemical characterization of plasma-activated water and plant material***

The chemical composition of plasma-activated water (PAW) and leaves was analysed spectrophotometrically using a UV–VIS spectrometer (Shimadzu UV-1900). Reactive nitrogen species (RNS) were quantified using commercially available assay kits according to the manufacturers' protocols. Nitrite ($NO_2^-$) concentration was determined by the Griess reagent (Cayman Chemical), with absorbance measured at 540 nm. Nitrate ($NO_3^-$) concentration was measured using a Spectroquant nitrate assay kit (Merck Chemicals), with absorbance recorded at 330 nm. Sulfamic acid was used to eliminate interference caused by nitrites during the nitrate determination. The concentration of hydrogen peroxide ($H_2O_2$) was determined based on its reaction with titanyl ions from titanium oxysulfate ($TiOSO_4$), forming a yellow peroxytitanium complex with an absorption maximum at 407 nm (Eisenberg, 1943). Sodium azide ($NaN_3$) was added to prevent the reaction of $H_2O_2$ with nitrite ions ($NO_2^-$) and to stabilise the samples following the plasma treatment. PAW composition was analysed daily over seven days after preparation, with measurements performed in at least three independent replicates. For the plant material analysis, reactive nitrogen species, primarily nitrite ($NO_2^-$) and nitrate ($NO_3^-$), were analysed following extraction of leaf tissue in 50 mM Na-phosphate buffer (pH 7.8). The extracts were subsequently analysed using the same spectrophotometric methods as described for PAW.

### *Plant material and experimental design*

Seeds of lettuce (*Lactuca sativa* L.), cultivar Matilda, were used in all experiments. Seeds were sown into inert mineral wool (rockwool) cubes placed in net pots inserted into openings in the lids of hydroponic cultivation containers and cultivated hydroponically for 10 weeks in a growth chamber under controlled environmental conditions. Plants were maintained in this system from sowing throughout the entire cultivation period without transplantation. The net pots were partially immersed in the hydroponic solution, while shoots developed above the container lids. Continuous aeration of the nutrient solution was provided using an air pump to ensure adequate oxygen supply to the root system (Figure 2). Plants were grown in a half-strength Hoagland nutrient solution with pH adjusted to 6.2 using HCl or KOH (Hoagland and Arnon, 1950) whereas the solution was changed every 7 days. The growth conditions were set as follows: temperature 23/25 °C (night/day), a 16-h photoperiod, and a photosynthetically active radiation (PAR) intensity of 250 μmol $m^{-2}$ $s^{-1}$.

### Experiment A: Standard nutrient conditions

Four types of treatments were established:

- control (**ctrl**) – plants cultivated in half-strength Hoagland solution,
- seed priming (**prim**) – seeds were soaked in plasma-activated water (PAW) for 1.5 h before sowing and subsequently cultivated in half-strength Hoagland solution,
- plasma-activated water cultivation (**PAW**) – plants cultivated in PAW mixed with salts of half-strength Hoagland solution,
- primed seeds and plasma-activated water cultivation (**prim PAW**) – seeds were soaked in PAW for 1.5 h before sowing, and plants were cultivated in PAW mixed with salts of half-strength Hoagland solution.

### Experiment B: Nitrogen-deprived conditions

The second experiment was conducted under the same growth conditions and experimental setup as in Experiment 1 but using a nitrogen-deprived half-strength

Hoagland solution. Nitrogen-containing macronutrients ($KNO_3$ and $Ca(NO_3)_2 \cdot 4H_2O$) were replaced by equimolar concentrations of KCl and $CaCl_2 \cdot 2H_2O$.
Four treatments were established again:

- nitrogen-deprived control (**ctrl**) – plants cultivated in half-strength nitrogen- deprived Hoagland solution,
- seed priming under nitrogen deprivation (**prim**) – seeds were soaked in PAW for 1.5 h before sowing and cultivated in nitrogen-deprived half-strength Hoagland solution,
- plasma-activated water cultivation under nitrogen deprivation (**PAW**) – plants cultivated in PAW supplemented with nitrogen-deprived half-strength Hoagland solution
- primed seeds and plasma-activated water cultivation under nitrogen deprivation (**prim PAW**) – seeds were soaked in PAW for 1.5 h before sowing, and plants were cultivated in PAW supplemented with salts to prepare a nitrogen-deprived half-strength Hoagland solution.

To compare the effect of cultivation, leaves at the same age were analysed in all subsequent analyses. Either young (the first two whorls) or old (the oldest two whorls) leaves were used for the enzymatic comparison, and the middle part of the lettuce rosettes (the second two whorls) was taken for pigments and non-enzymatic antioxidants analysis.

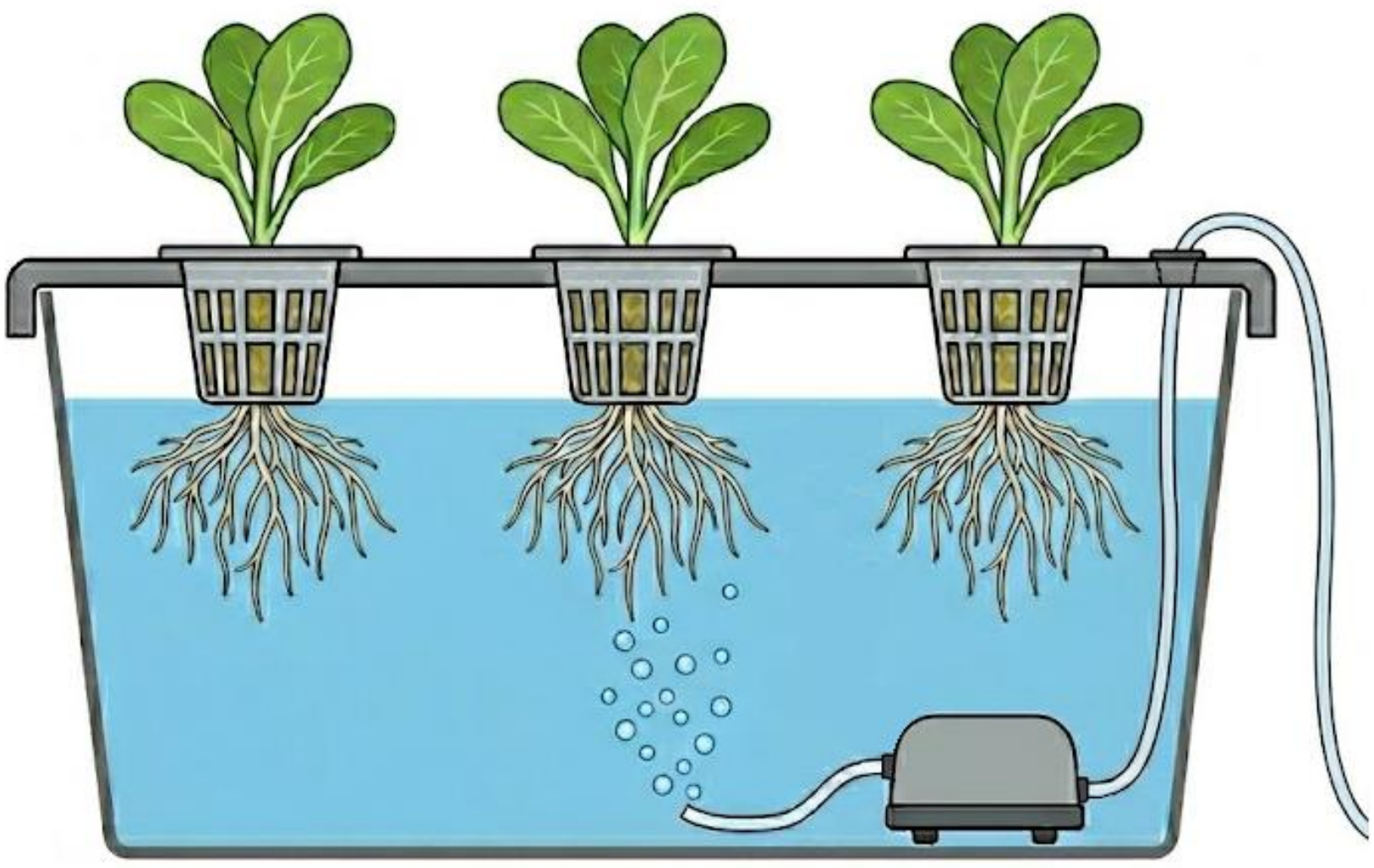

*Fig. 2 Schematic diagram of the hydroponic cultivation setup with continuous aeration of the nutrient solution.*

### *Determination of growth parameters*

When all plants had developed at least two leaf whorls, corresponding to the 4$^{th}$ week of cultivation, digital photographs of lettuce heads were taken weekly. Images were analysed using ImageJ software (NIH, USA) to determine the projected area of the lettuce head. After 10 weeks of cultivation, the fresh weight (FW) of lettuce heads was measured. Subsequently, plant material was dried in a drying oven at 70 °C for 4 days, after which the dry weight (DW) was determined.

***Determination of photosynthetic pigments and their degradation products***

Photosynthetic pigments (chlorophyll *a* and *b*) as well as accessory photosynthetic pigments, carotenoids, with their degradation products (pheophytins) were determined according to the method of (Hartmut K, 1987) and Lichtenthaler and Buschmann (Lichtenthaler and Buschmann, 2001).The middle whorls of lettuce heads (leaves 3-4) were analysed. Pigments were extracted from the fresh leaves with 80% (v/v) acetone. After extraction, the samples were centrifuged (Centrifuge Mega Star 1.6R, VWR, Osterode am Harz, Germany) at room temperature for 5 min at 500 x g so the extract would be fully clear, and the absorbance of the supernatant was measured spectrophotometrically using a UV–VIS spectrophotometer (Jenway 6705 UV/Vis, Jenway, London, UK). Absorbances were recorded at 663.2 nm, 646.8 nm, and 470 nm, and pigments were calculated with the equations where V present the volume of the extraction solution:

$$chlorophyll\ a+b\ = \frac{(7.15\ A_{663.2}-18.71 A_{646.8})\ x\ V}{60}\ \text{mg g}^{-1}\ \text{FW}$$

$$pheophytin\ a+b\ = \frac{(3.84\ A_{663.2}+33.6 A_{646.8}\ )\ x\ V}{60}\ \text{mg g}^{-1}\ \text{FW}$$

$$carotenoids = \frac{\frac{1000 A_{470}\ -1.82 C_{ch\ a}\ -85.02 C_{chl\ b}}{198}}{60}\ \text{mg g}^{-1}\ \text{FW}$$

**Determination of antioxidant enzymes**

**Protein extraction and determination**

For the determination of soluble protein content and antioxidant enzyme activities, young and old leaves of rosettes were separately compared. Leaf samples were harvested, immediately frozen in liquid nitrogen, and ground to a fine powder using a mortar and pestle. The resulting powder was homogenized in 50 mM sodium phosphate buffer (pH 7.8) containing 1 mM EDTA, a protease inhibitor cocktail, 5 mM ascorbate and 2% polyvinylpolypyrrolidone. The homogenates were centrifuged twice: first at 2,000 × g for 5 min, followed by a second centrifugation at 13,000 × g for 20 min (Centrifuge Mega Star 1.6R, VWR, Osterode am Harz, Germany). The resulting supernatant was collected and used for soluble protein determination (Bradford, 1976) and for subsequent antioxidant enzyme activity assays. Absorbance was measured at 595 nm with a UV–VIS spectrophotometer (Jenway 6705 UV/Vis, Bibby Scientific Ltd., Essex, UK).

**Activity of ascorbate peroxidase (APX)**

Ascorbate peroxidase (APX; EC 1.11.1.11) activity was determined spectrophotometrically according to Nakano and Asada (Nakano and Asada, 1981) by monitoring the decrease in absorbance at 290 nm resulting from ascorbate (ASC) oxidation. Enzyme activity was calculated following the protocol by (Bogomolets, 1985)using the molar extinction coefficient for ascorbate (ε = 2.8 $mM^{-1}$ $cm^{-1}$) and expressed as nM of ASC oxidized per min per μg of soluble proteins.

**Activity of glutathione reductase (GR)**

Glutathione reductase (GR; EC 1.8.1.7) activity was determined spectrophotometrically according to Foyer and Halliwell (Foyer and Halliwell, 1976) by monitoring the oxidation of NADPH as a decrease in absorbance at 340 nm. Enzyme activity was calculated following (Bogomolets, 1985) using the molar extinction coefficient for NADPH (ε = 6.2 $mM^{-1}$ $cm^{-1}$) and expressed as μM of NADPH oxidized per min per mg of soluble proteins.

**Activity of guaiacol peroxidase (G-POX)**

Guaiacol peroxidase (G-POX; EC 1.11.1.7) activity was assayed spectrophotometrically according to Frič and Fuchs (Frič and Fuchs, 1970) by measuring the formation of tetraguaiacol from guaiacol in the presence of $H_2O_2$. The increase in absorbance was recorded at 440 nm. Enzyme activity was calculated according to Chance and Maehly (Chance and Maehly, 1955) using the molar extinction coefficient for tetraguaiacol (ε = 26.6 $mM^{-1}$ $cm^{-1}$) and expressed as μM of tetraguaiacol formed per min per μg of soluble proteins.

**Activity of superoxide dismutase (SOD)**

Superoxide dismutase (SOD; EC 1.15.1.1) activity was determined according to (García-Limones et al., 2002) by monitoring the inhibition of the photochemical reduction of nitro blue tetrazolium (NBT; Merck). The reaction was initiated by exposing the reaction mixture to cool-white light at an intensity of 50 μmol $m^{-2}$ $s^{-1}$ for 20 min. Non-illuminated samples and illuminated reaction mixtures without enzyme extract served as blanks and calibration controls, respectively. Absorbance of the reaction products was measured spectrophotometrically at 560 nm. One unit of SOD activity was defined as the amount of enzyme required to cause 50% inhibition of NBT photoreduction under the assay conditions per mg of soluble proteins.

**Determination of non-enzymatic antioxidants**

**Determination of soluble phenolics**

Soluble phenolic content was determined from the middle leaves (leaves 3–4) using the Folin–Ciocalteu reagent, following Ainsworth and Gillespie (Ainsworth and Gillespie, 2007) protocol. Absorbance was measured at 765 nm, and phenolic concentration was calculated from a gallic acid calibration curve (0–500 μg $mL^{-1}$) and expressed as gallic acid equivalents (GAE) in milligrams per gram of fresh weight (FW).

**Determination of quercetin**

Quercetin was determined using the aluminium chloride (AlCl3) colorimetric method, as described by Zivcak (Zivcak et al., 2017). Absorbance was measured at 415 nm, and the concentration was calculated from a quercetin calibration curve (25–100 μg $mL^{-1}$), expressed as quercetin equivalents (QE) in milligrams per gram of fresh weight (FW).

**Sensory evaluation**

Lettuce samples were served on plates marked with numeric codes corresponding to each variant to prevent bias. The sensory panel consisted of untrained volunteers (n = 15), including students and staff of the Faculty of Natural Sciences in Bratislava (both men and women, aged 23–72 years), who evaluated the samples using a 10-point intensity scale (1 = lowest, 10 = highest). Sensory attributes assessed included appearance, aroma, flavour, texture, and overall perception. Each participant recorded their responses on a short anonymous form. Sensory evaluation was performed only for plants cultivated under nitrogen-sufficient conditions. No exclusion criteria were applied for the analysis. As this assessment was exploratory in nature, no sample size calculation or statistical power analysis was performed. The panel size was determined by the availability of plant material within the integrated experimental design.

**Ethics statement**

This study involving human participants was conducted in accordance with the ethical principles of the Declaration of Helsinki and relevant institutional guidelines. Ethical approval was obtained from the Ethics Committee of Comenius University, Bratislava, Slovakia, on April 30, 2026. All participants provided informed consent prior to participation in the sensory evaluation. Participation was voluntary, and participants were informed of their right to withdraw at any time without consequences. Privacy rights of the participants were respected, and all collected data were anonymized prior to analysis.

**Statistical analysis**

The results were statistically processed and evaluated by Microsoft Excel (Office 365) and STATGRAPHICS Centurion XV (Statgraphics Technologies, version 15.2.05). All experiments and analyses were repeated at least three times with five to seven plants per treatment per replication. Based on statistical analyses (one-way, two- and three-factor ANOVA, with post-hoc Least Significant Difference, LSD test), we evaluated the significance of differences in the results between treatments and factors at the 5% level of significance. Principal Component Analysis (PCA) was employed to identify patterns, associations of variables and major sources of variability within the data sets.

## Results

**RONS evaluation**

Analysis of the most important RONS in the plasma-activated water and in its combinations with the half-strength Hoagland solution (N-supplemented and N-deprived) used for hydroponic lettuce cultivation showed significant differences (Fig. 3). As expected, the amount of $NO_3^-$ in the N-deprived Hoagland solution was very low; however, after activation with PAW, its concentration increased more than 40 times and remained stable during the 7-day measurement. Almost no change was detected in N-supplemented Hoagland solution after its activation with PAW; in both cases, the concentration of $NO_3^-$ was very stable and reached approximately 27 mM. A different situation was observed in a case of $NO_2^-$. They were in very low concentrations in both Hoagland solutions, independent of the N-supplementation; however, more than 1 mM was measured after PAW activation. Hydrogen peroxide proved to be the least stable molecule, with concentrations under the measurement limit within both Hoagland solutions without activation. After PAW activation, in both cases, its concentration reached up to 260 μM but continuously decreased within days post activation, more drastically in the N-deprived Hoagland solution, where its content was near zero from the 4th day of measurement. On the other hand, there was still about 17 μM of $H_2O_2$ on the 7th day of measurement in N-supplemented PAW-activated Hoagland.

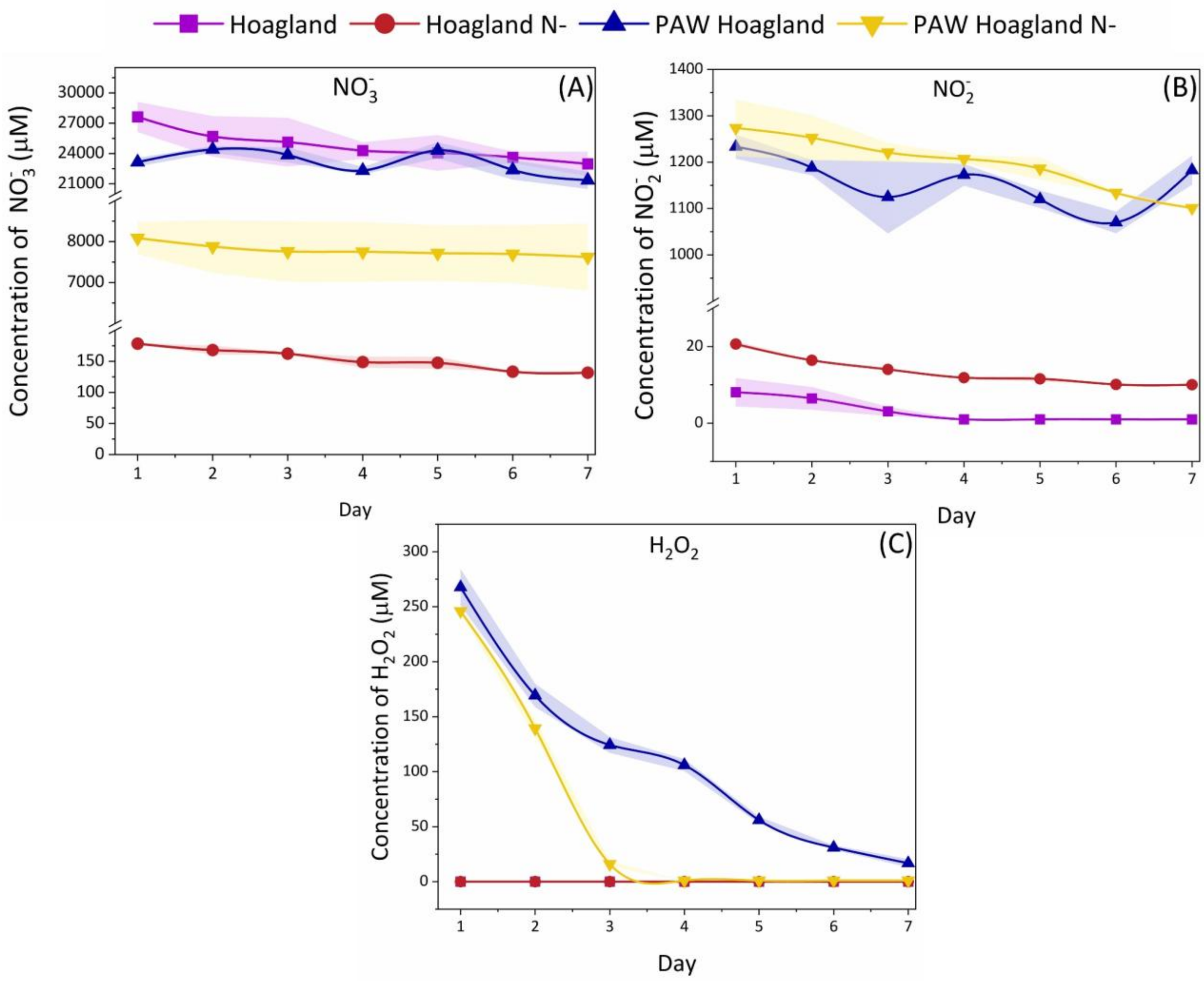

*Fig. 3. Concentration of $NO_3$ (A)$^-$, $NO_2^-$ (B) and $H_2O_2$ (C) in the half-strength Hoagland solution (purple line), in PAW mixed with salts of the half-strength Hoagland solution (blue line), in the N-deprived (N-) half-strength Hoagland solution (red line) and in PAW mixed with salts of the N-deprived (N-) half-strength Hoagland solution (yellow line). Values represent means and shaded areas represent the standard deviation (SD) of the measurements.*

**Growth parameters**

Growth of the lettuce aboveground parts (heads) was significantly affected by both N-supplementation and PAW treatments (Figs. 4, 5, 6; Tab.1). The head area increased significantly from week to week across all treatments (Tab. 1). Nitrogen deprivation caused a pronounced reduction in the lettuce growth (Fig. 5B). While seed priming in the N-deprived solution (prim) did not improve growth, cultivation in the PAW-activated Hoagland (PAW) or the combination of seed priming and the PAW-activated Hoagland (prim PAW) resulted in a significant growth enhancement. Among the N-deprived treatments, PAW was the most effective in promoting the lettuce growth (Tab. 1B).

In contrast, lettuces grown in N-supplemented solutions exhibited the highest growth under the prim PAW treatment (Fig. 5A; 4; Tab. 1A). However, unlike in the N-deprived conditions, the PAW treatment alone resulted in the lowest growth under N-sufficient conditions.

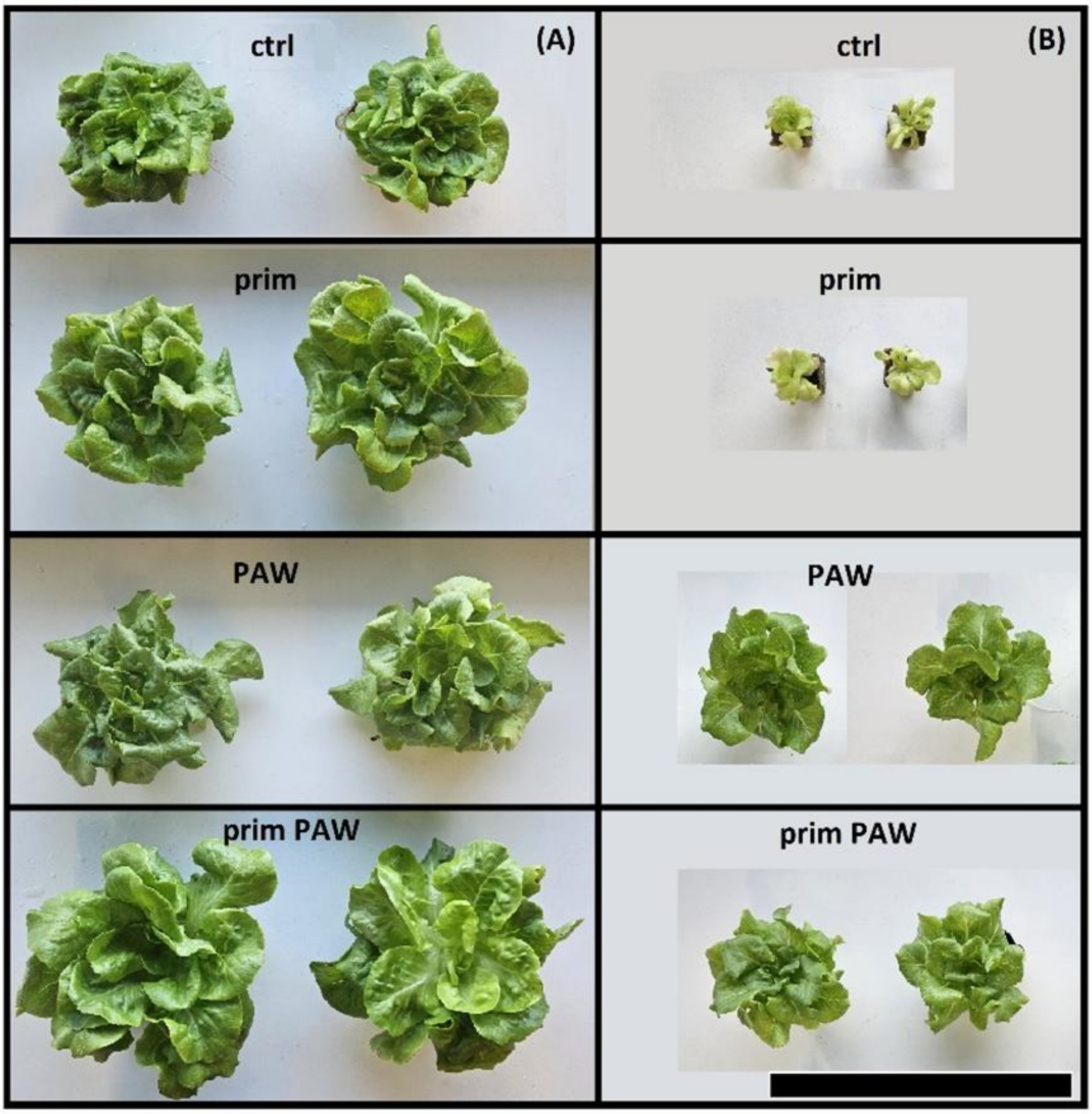

*Fig. 4 Overall appearance of lettuce plants after 10 weeks of cultivation in the half-strength Hoagland solution in N-supplemented (A) and N-deprived (B) conditions in combination with various PAW treatments: seed priming (prim), cultivation in PAW-activated Hoagland solution (PAW), and seed priming combined with cultivation in PAW-activated Hoagland solution (prim PAW). Scale bar = 30 cm.*

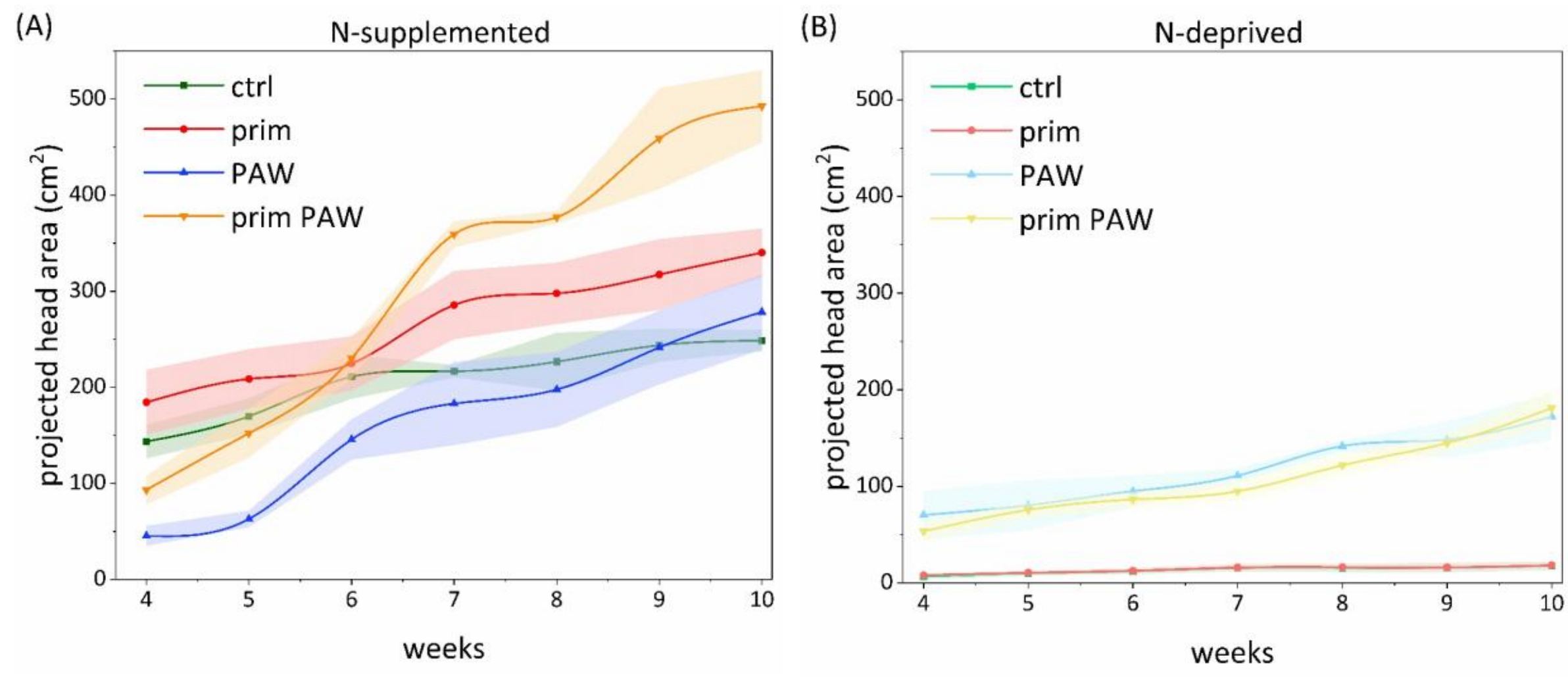

*Fig. 5 Projected lettuce head area during the 10 weeks of cultivation in half-strength Hoagland solution in N-supplemented (A) and N-deprived (B) solution in combination with various PAW treatments: seed priming (prim), cultivation in PAW-activated Hoagland solution (PAW), and seed priming combined with cultivation in PAW-activated Hoagland solution (prim PAW). Values represent means and shaded areas represent the standard deviation (SD) of the measurements.*

*Tab. 1 Two-factor ANOVA for the lettuce head area in Hoagland solution with (A) or without (B) N-supplementation with LSD test showing the differences between the tested groups.*

| (A) FACTOR IN N-SUFFICIENT CONDITIONS | HEAD AREA | LSD TEST FOR X | HOMOGEN. GROUPS | LSD TEST FOR Y | HOMOGEN. GROUPS |
|---|---|---|---|---|---|
| TREATMENT (X) | *** | PAW | a | week 4 | a |
| WEEK (Y) | *** | ctrl+ | b | week 5 | b |
| X x Y | *** | prim | c | week 6 | c |
| | | prim PAW | d | week 7 | d |
| | | | | week 8 | e |
| | | | | week 9 | f |
| | | | | week 10 | g |
| **(B) FACTOR IN N-DEPRIVED CONDITIONS** | | | | | |
| TREATMENT (X) | *** | ctrl | a | week 4 | a |
| WEEK (Y) | *** | prim | a | week 5 | b |
| X x Y | *** | prim PAW | b | week 6 | c |
| | | PAW | c | week 7 | d |
| | | | | week 8 | e |
| | | | | week 9 | f |
| | | | | week 10 | g |

*** $p < 0.001$

Fresh weight increased 1.4-times in prim PAW under N-supplementation (Fig. 6A); however, more than 18-times in prim PAW under N-deprivation (Fig. 6B) in comparison with the corresponding controls. Of course, this 18-fold increase should be taken in the context of very low biomass of the control plants under N-deprivation. Results for the percentage of dry weight (% DW) were less pronounced than those observed for FW; all treated plants accumulated comparable amounts of dry matter relative to FW. This indicated that the significant differences observed in head area and FW among treatments were primarily driven by changes in water accumulation rather than alterations in biomass production (Fig. 6). Accordingly, variations in lettuce growth across treatments were more closely associated with water management and tissue hydration status, rather than with dry matter allocation.

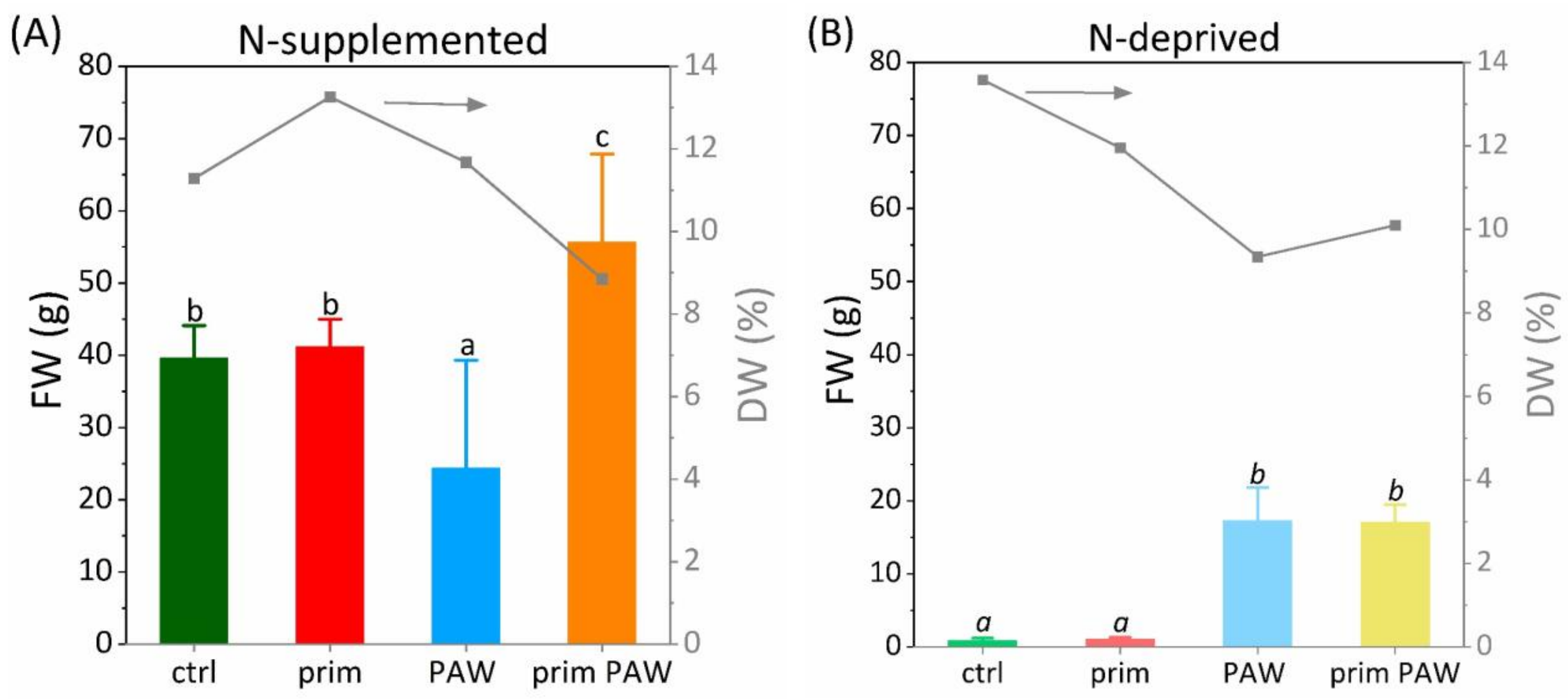

*Fig. 6 Fresh weight (FW) and percentage of dry weight (% DW) of lettuce cultivated in the half-strength Hoagland solution in N-supplemented (A) and N-deprived (B) solution in combination with various PAW treatments: seed priming (prim), cultivation in PAW-activated Hoagland solution (PAW), and seed priming combined with cultivation in PAW-activated Hoagland solution (prim PAW). Values represent means ± SD. Different letters indicate significant differences at p ≤ 0.05 (LSD test).*

## Photosynthetic pigments

Nitrogen availability significantly affected the content of major and accessory photosynthetic pigments (Figs. 7, 9; Tab. 2), as well as chlorophyll degradation which closely followed the changes in chlorophyll concentrations (Fig. 8). However, cultivation in PAW-activated Hoagland solution (PAW) and in the combined prim PAW treatments significantly increased the chlorophyll levels compared with the control and prim treatment under N-deprived conditions (Fig. 7B).

*Tab. 2 Two-factor ANOVA (for pigments and metabolites) and three-factor ANOVA (for enzymes) in lettuce plants with LSD test showing the differences between the tested groups.*

| ***FACTOR*** | **CHL *a+b*** | **PHEO. *a+b*** | **CAROT.** | **APX** | **GR** | **G-POX** | **SOD** | **PHEN.** | **QUERCETIN** |
|---|---|---|---|---|---|---|---|---|---|
| **LEAVES (X)** | | | | *** | ns. | *** | *** | | |
| **NITROGEN (Y)** | *** | *** | *** | ns. | ns. | *** | *** | ns. | *** |
| **TREATMENT (Z)** | *** | *** | *** | ** | * | *** | ** | ns. | *** |
| X x Y | | | | ns. | *** | *** | ns. | | |
| X x Z | | | | *** | ns. | *** | ns. | | |
| Y x Z | *** | *** | *** | ** | ** | ns. | * | | *** |
| X x Y x Z | | | | ns. | *** | *** | * | | ns. |
| ***LSD TEST FOR "X"*** | | | | | | | | | |
| YOUNG | | | | b | | a | a | | |
| OLD | | | | a | | b | b | | |
| **LSD TEST FOR "Y"** | | | | | | | | | |
| N SUPPLEMENT. | b | b | b | | | b | a | | b |
| N$^-$ SUPPLEMENT. | a | a | a | | | a | b | | a |
| **LSD TEST FOR "Z"** | | | | | | | | | |
| CTRL | a | a | b | ab | a | a | b | | a |
| PRIM | a | a | a | a | ab | b | a | | b |
| PAW | b | b | b | b | b | c | a | | b |
| PRIM PAW | b | b | c | b | b | bc | a | | b |

*ns. nonsignificant effect; * p < 0.05; *** p < 0.01 *** p < 0.001*

Under N-supplemented conditions, both PAW and prim PAW resulted in significantly higher chlorophyll concentrations relative to the control, whereas seed priming alone had no significant effect (Fig. 7A; Tab. 2). Chlorophyll concentration was affected negatively in N-deprived solution (Fig. 7B), however, after PAW activation, in PAW and prim PAW-treated plants, this effect was alleviated, and plants achieved significantly more photosynthetic pigments than control plants and plants only primed by PAW (prim treatment). Chlorophyll *a+b* increased up to 1.3-times under PAW and prim PAW treatments (Fig. 7A), however, up to 3.2-times under PAW and prim PAW treatments under N-deprivation (Fig. 7B) in comparison with the corresponding controls. However, it is important to note that the amount of chlorophyll did not reach the levels observed in the respective N-supplemented treatments.

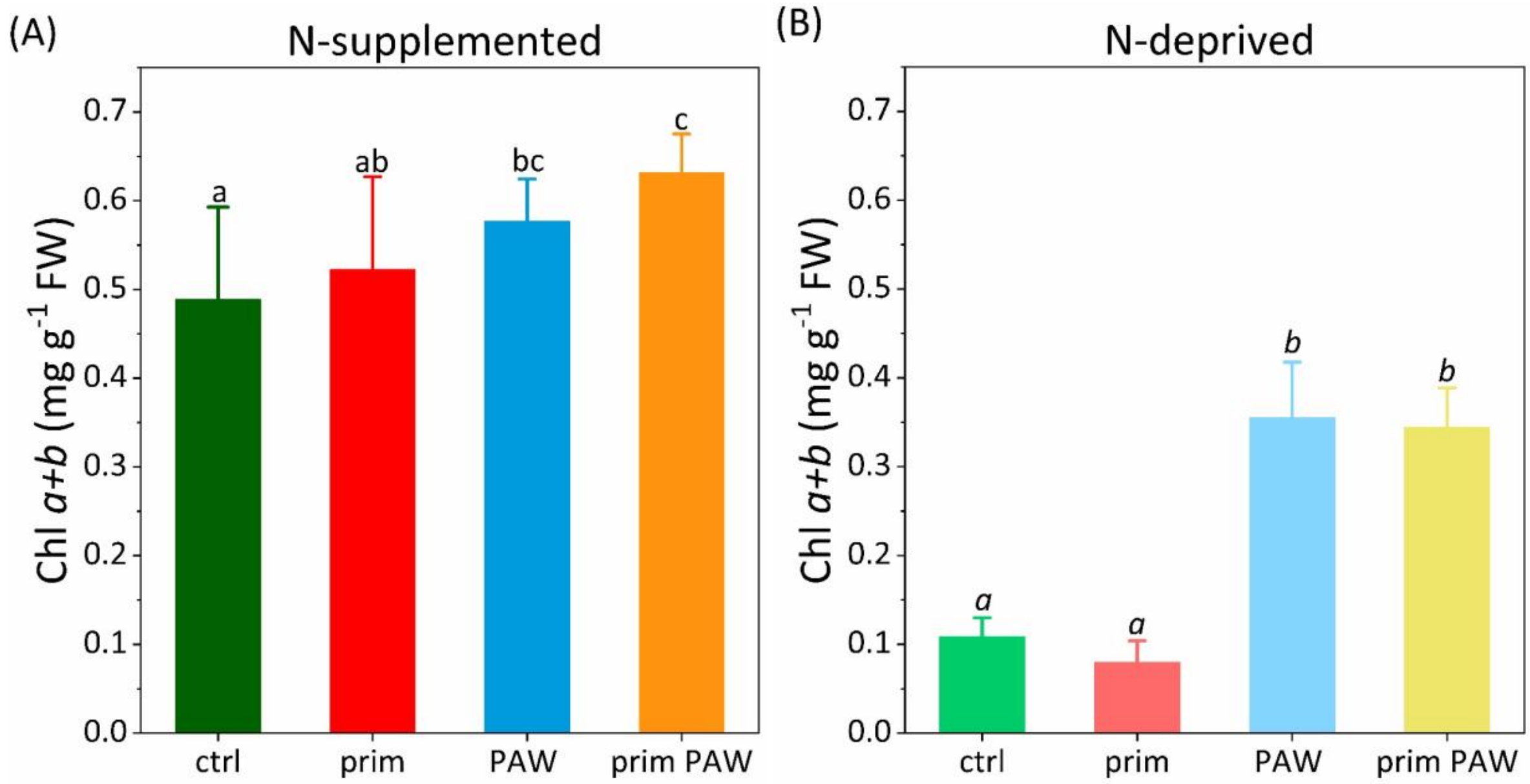

*Fig. 7 Total chlorophyll (chl a+b) of lettuce cultivated in the half-strength Hoagland solution under N-supplemented (A) and N-deprived (B) conditions in combination with different PAW treatments: seed priming (prim), cultivation in PAW-activated Hoagland solution (PAW), and seed priming combined with cultivation in PAW-activated Hoagland solution (prim PAW). Values represent means ± SD. Different letters indicate significant differences at $p \leq 0.05$ (LSD test).*

The concentration of pheophytin *a+b* reflected the trends observed for chlorophyll content (Fig. 8), with similar trends confirmed by the two-way ANOVA (Tab. 2). Under N-supplemented conditions, PAW-based treatments maintained higher pheophytin levels relative to the control, the increase was up to 1.2-times (Fig. 8A). Nitrogen deprivation led to a pronounced reduction in pheophytin levels, indicating overall pigment depletion (Fig. 8B). In these conditions, PAW and prim PAW treatments significantly increased pheophytin concentrations compared with the control (up to 3-times), consistent with their positive effect on chlorophyll accumulation.

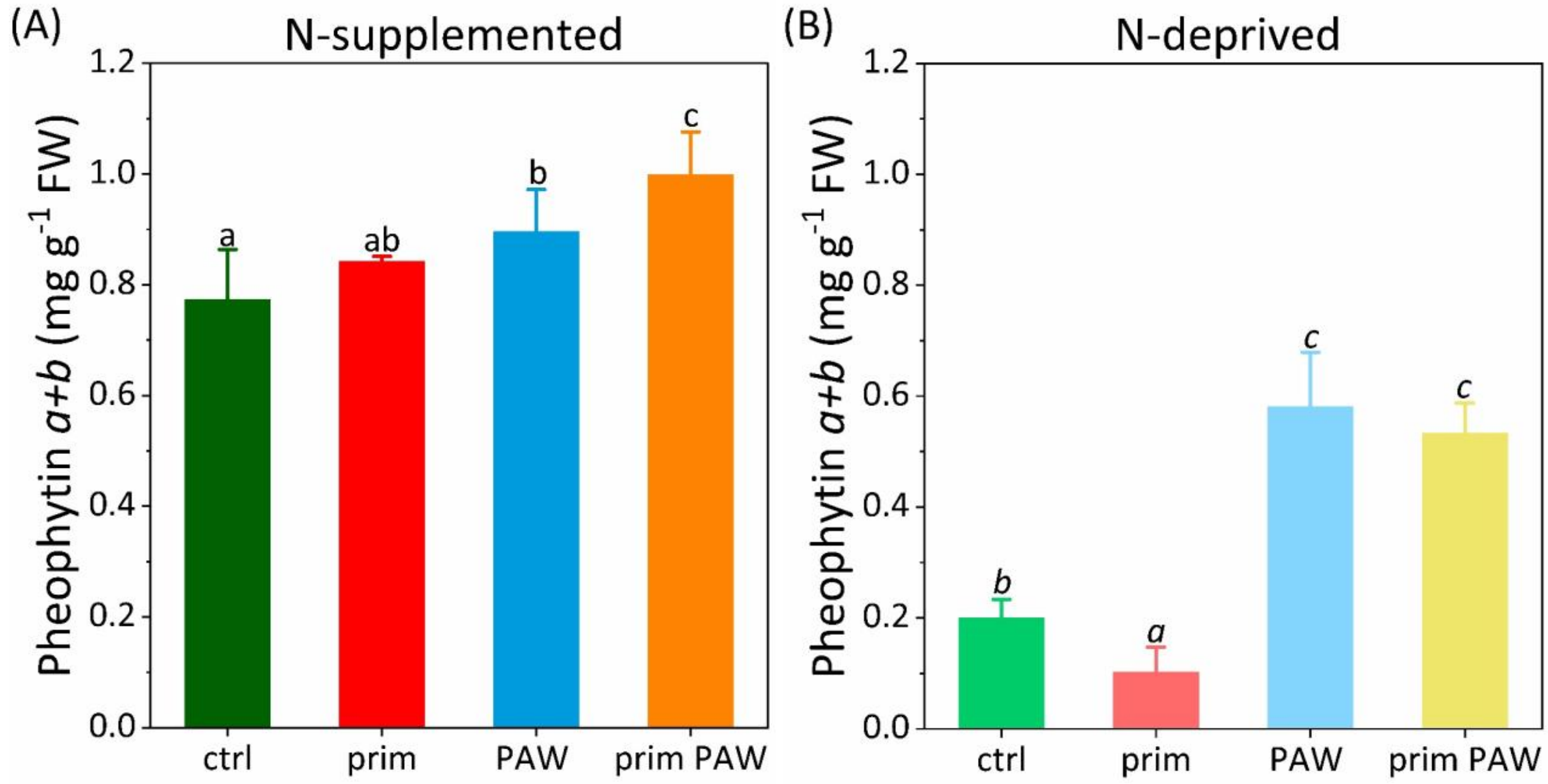

*Fig. 8 Concentration of pheophytin a+b of lettuce cultivated in the half-strength Hoagland solution under N-supplemented (A) and N-deprived (B) conditions in combination with different PAW treatments: seed priming (prim), cultivation in PAW-activated Hoagland solution (PAW), and seed priming combined with cultivation in PAW-activated Hoagland solution (prim PAW). Values represent means ± SD. Different letters indicate significant differences at p ≤ 0.05 (LSD test).*

Accessory photosynthetic pigments and non-enzymatic antioxidants – carotenoids, were very significantly influenced by nitrogen supplementation and PAW treatments (Fig. 9; Tab. 2). Under N-supplemented conditions, PAW and prim PAW treatments led to a significant decrease in carotenoid concentration compared with the control and prim treatment (Fig. 9A), indicating a differential response of carotenoid metabolism to PAW depending on nitrogen availability. In contrast, N-deprivation resulted in a marked decrease in carotenoid concentrations in control and prim treatments. This reduction was significantly alleviated by PAW and prim PAW treatments, where carotenoids increased up to 1.7-times in comparison to the corresponding control (Fig. 9B).

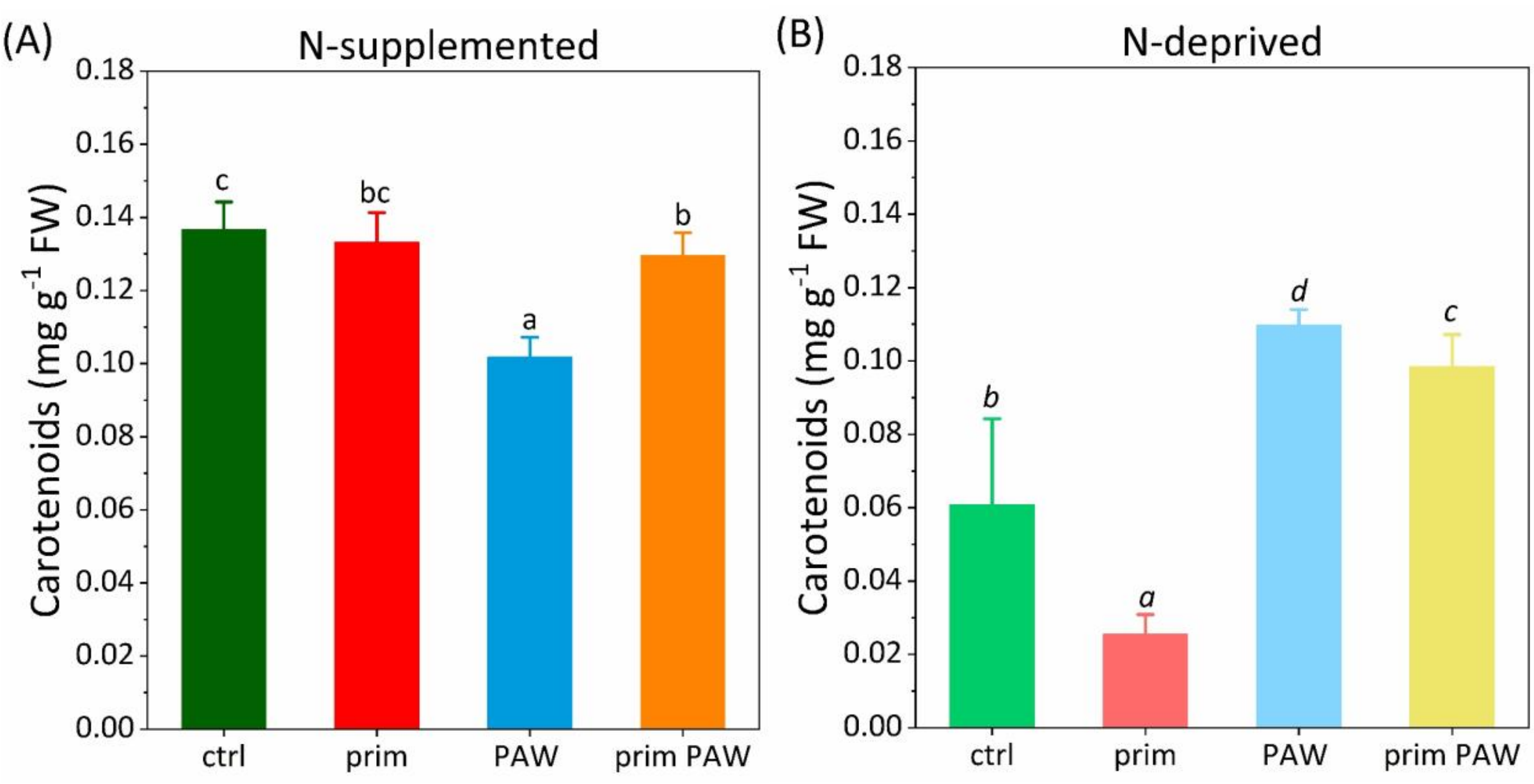

*Fig. 9 Concentration of carotenoids of lettuce cultivated in the half-strength Hoagland solution under N-supplemented (A) and N-deprived (B) conditions in combination with different PAW treatments: seed priming (prim), cultivation in PAW-activated Hoagland solution (PAW), and seed priming combined with cultivation in PAW-activated Hoagland solution (prim PAW). Values represent means ± SD. Different letters indicate significant differences at p ≤ 0.05 (LSD test).*

**Antioxidant enzymes and secondary metabolites**

Ascorbate peroxidase (APX) achieved higher activities in young leaves than in old leaves (Fig. 10; Tab. 2). Nitrogen supplementation or deprivation was not a significant factor for APX; there were no differences in APX activity when comparing N-supplemented with N-deprived controls (Tab. 2). The highest activities were mostly observed in PAW and prim PAW treatment in young leaves, with an increase of approximately 2.2- and 2.5-times, respectively, in comparison with the corresponding controls (Fig. 10A, B). An increase in APX activity in old leaves was only observed in prim under N-supplemented conditions and in PAW (2.3-times higher than the corresponding control) and prim PAW treatments in N-deprived conditions in comparison to controls (Fig. 10C, D).

Young leaves

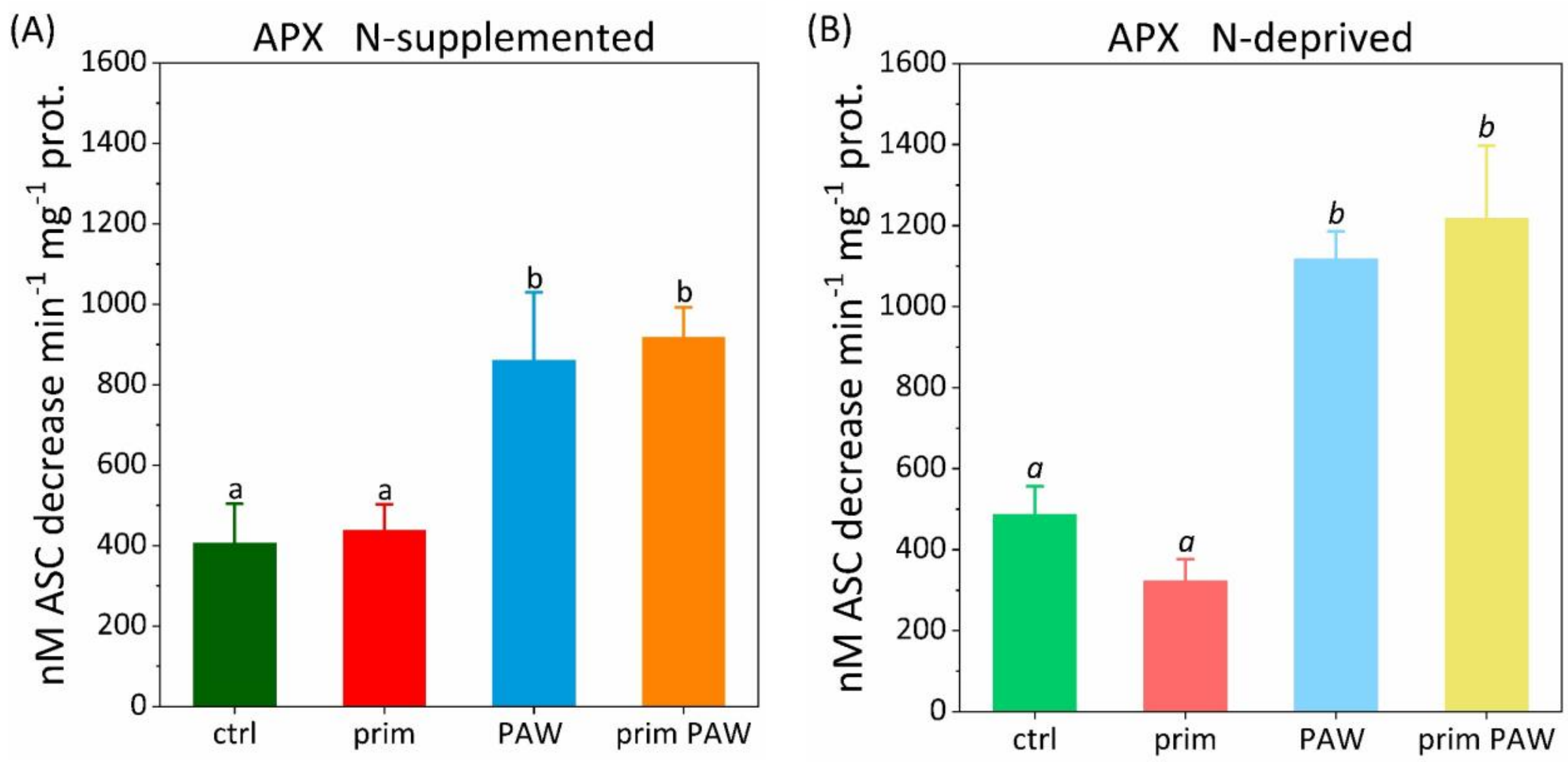

Old leaves

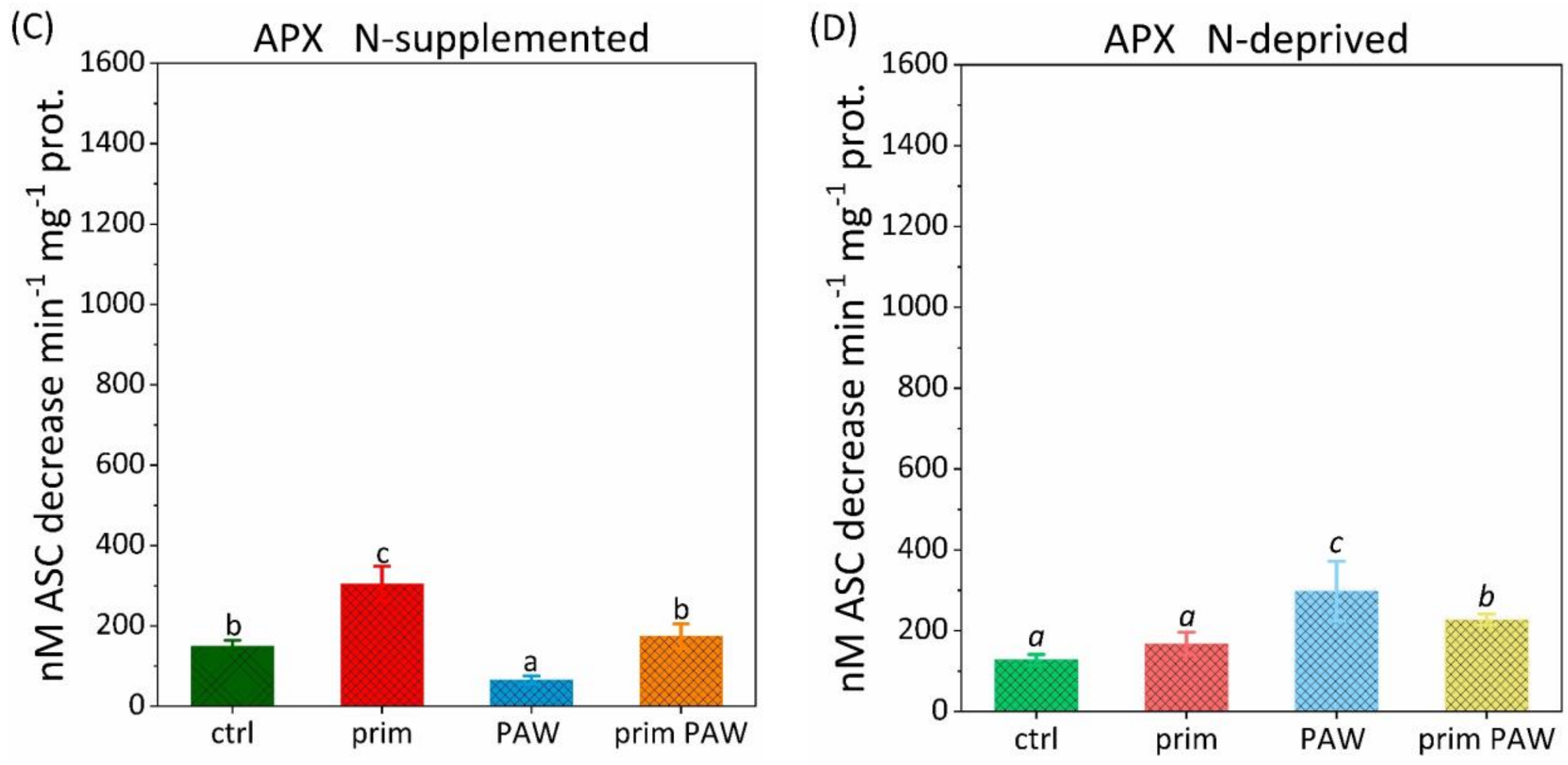

*Fig. 10 Activity of ascorbate peroxidase (APX) in young (A, B) and old (C, D) leaves of lettuce cultivated in the half-strength Hoagland solution under N-supplemented (A, C) and N-deprived (B, D) conditions in combination with different PAW treatments: seed priming (prim), cultivation in PAW-activated Hoagland solution (PAW), and seed priming combined with cultivation in PAW-activated Hoagland solution (prim PAW). Values represent means ± SD. Different letters indicate significant differences at $p \leq 0.05$ (LSD test).*

The only main factor with a significant impact on glutathione reductase (GR) activity was the factor treatment (Z, Tab. 2). Nitrogen deprivation significantly reduced GR activity in young leaves (Fig. 11A, B), and the opposite effect was observed in old leaves (Fig. 11C, D). Priming treatment in leaves in N-supplemented media led to a decrease in the GR activity; on the contrary, prim PAW treatment increased it approximately 1.3-times in young leaves (Fig. 11A). In old leaves in the N-deprived solution, prim, PAW, and prim PAW-treated plants showed significantly higher activity than the corresponding control; the increases in prim, PAW and prim PAW treatments were more than 2-times (Fig. 11D), but the opposite effect was achieved in N-deprived young leaves (Fig. 11B).

Young leaves

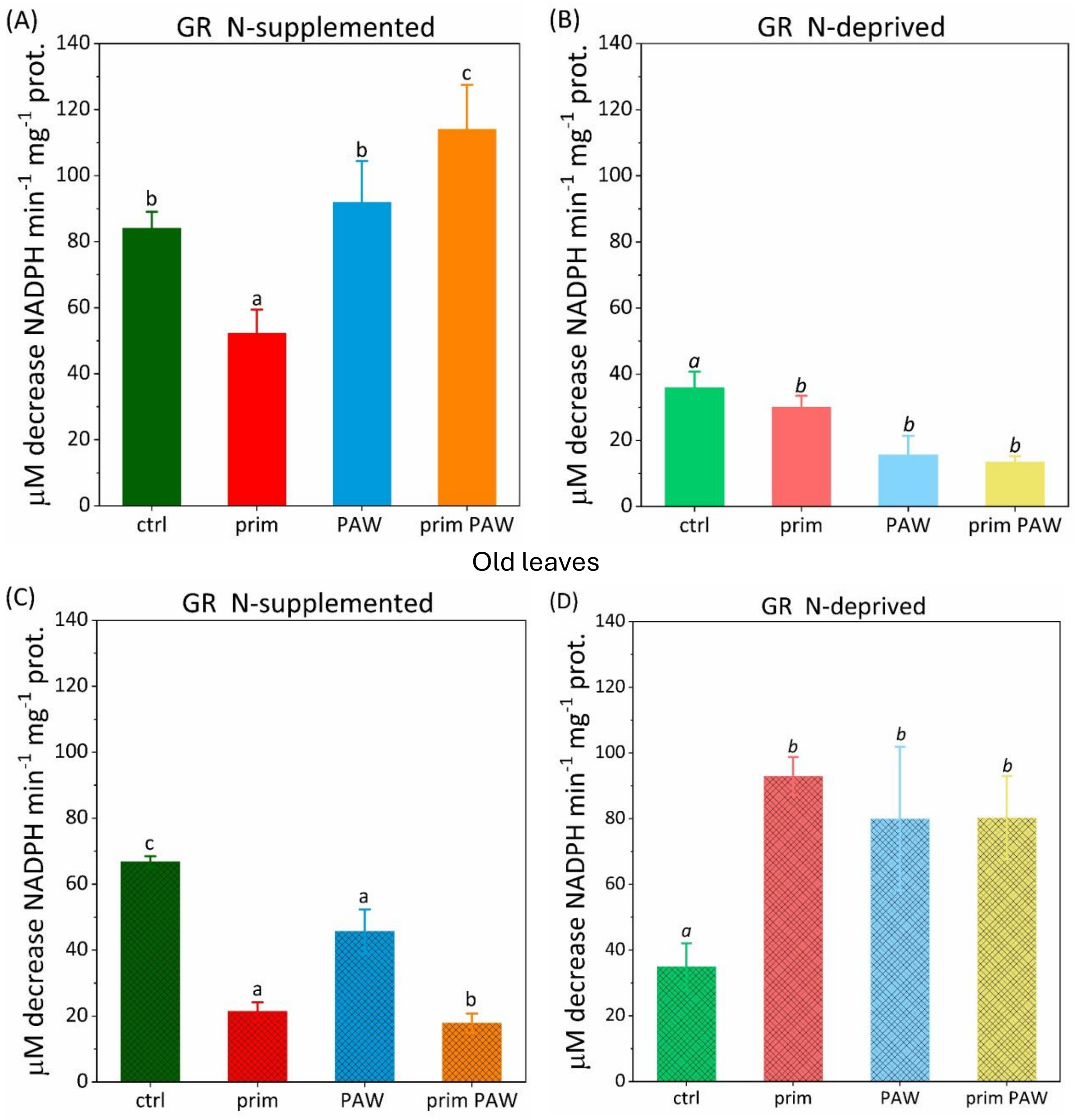

*Fig. 11 Activity of glutathione reductase (GR) in young (A, B) and old (C, D) leaves of lettuce cultivated in the half-strength Hoagland solution under N-supplemented (A, C) and N-deprived (B, D) conditions in combination with different PAW treatments: seed priming (prim), cultivation in PAW-activated Hoagland solution (PAW), and seed priming combined with cultivation in PAW-activated Hoagland solution (prim PAW). Values represent means ± SD. Different letters indicate significant differences at $p \leq 0.05$ (LSD test).*

All three selected main factors (X-leaves, Y-nitrogen supplementation, and Z-treatment) significantly influenced the G-POX activity after the three-way ANOVA (Tab. 2). The old leaves in N-supplemented solutions were significantly more active than the young leaves (Fig. 12A, C; Tab. 2). No significant increase was achieved in N-supplemented media in young leaves (Fig. 12A), however, in all plants treated with PAW (prim, PAW, prim PAW), a significant increase (1.2 to 1.4-times) in G-POX activity was observed in old leaves (Fig. 12C), with also much higher G-POX in the control. Similar increase in G-POX activity was confirmed in almost all treatments with PAW under N-deprivation (Fig. 12B, D) with an increase of 1.4 to 1.8-times in comparison with the corresponding controls.

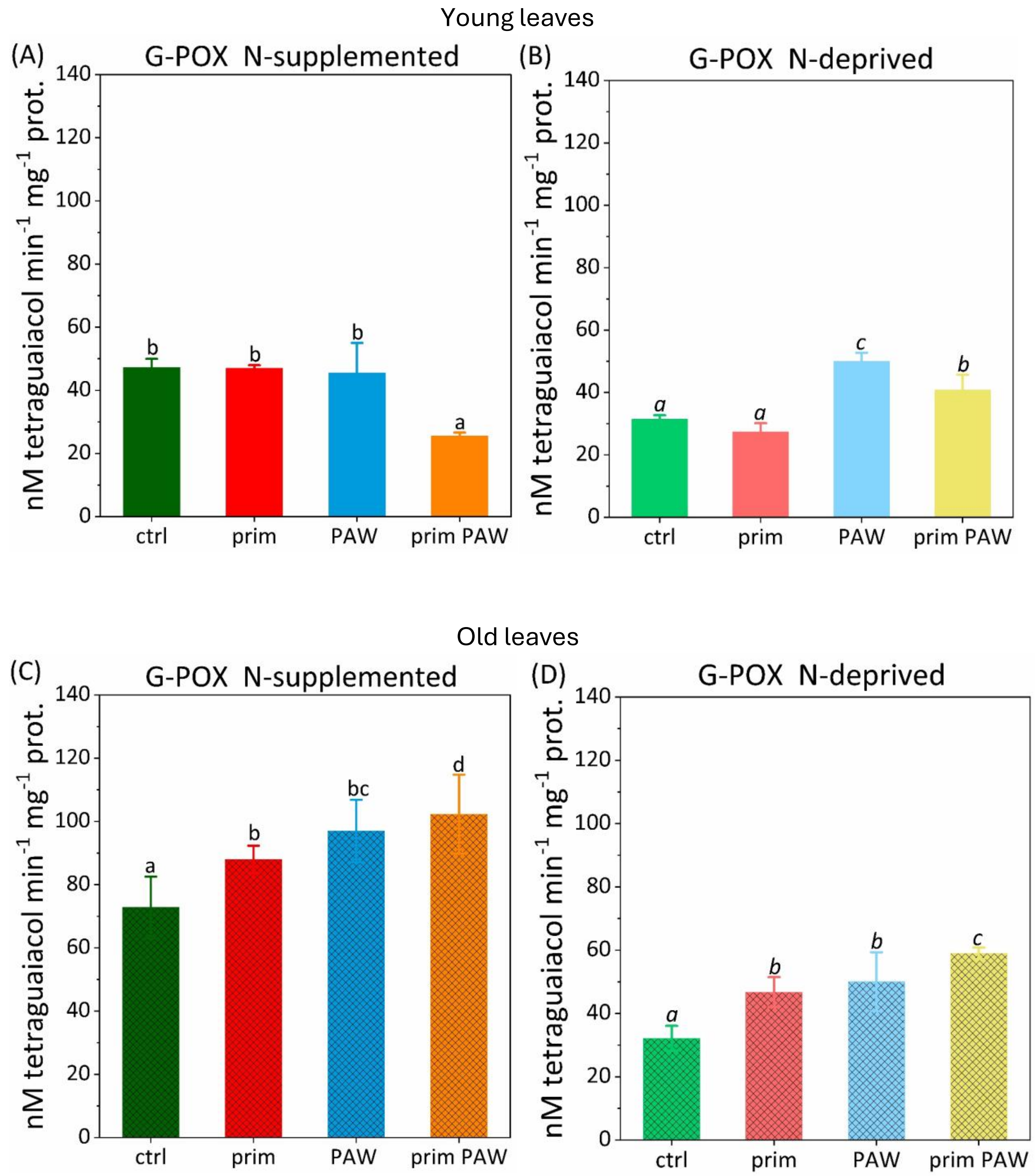

*Fig. 12 Activity of guaiacol peroxidase (G-POX) in young (A, B) and old (C, D) leaves of lettuce cultivated in the half-strength Hoagland solution under N-supplemented (A, C) and N-deprived (B, D) conditions in combination with different PAW treatments: seed priming (prim), cultivation in PAW-activated Hoagland solution (PAW), and seed priming combined with cultivation in PAW-activated Hoagland solution (prim PAW). Values represent means ± SD. Different letters indicate significant differences at $p \leq 0.05$ (LSD test).*

All three selected factors (X-leaves, Y-nitrogen supplementation, and Z-treatment) also influenced the superoxide dismutase (SOD) activities, with higher levels observed in old leaves and under N-deprived conditions (Tab. 2). Overall, PAW-treated plants across all treatments showed no increase but mostly a decrease in SOD activity compared to controls (Fig. 13).

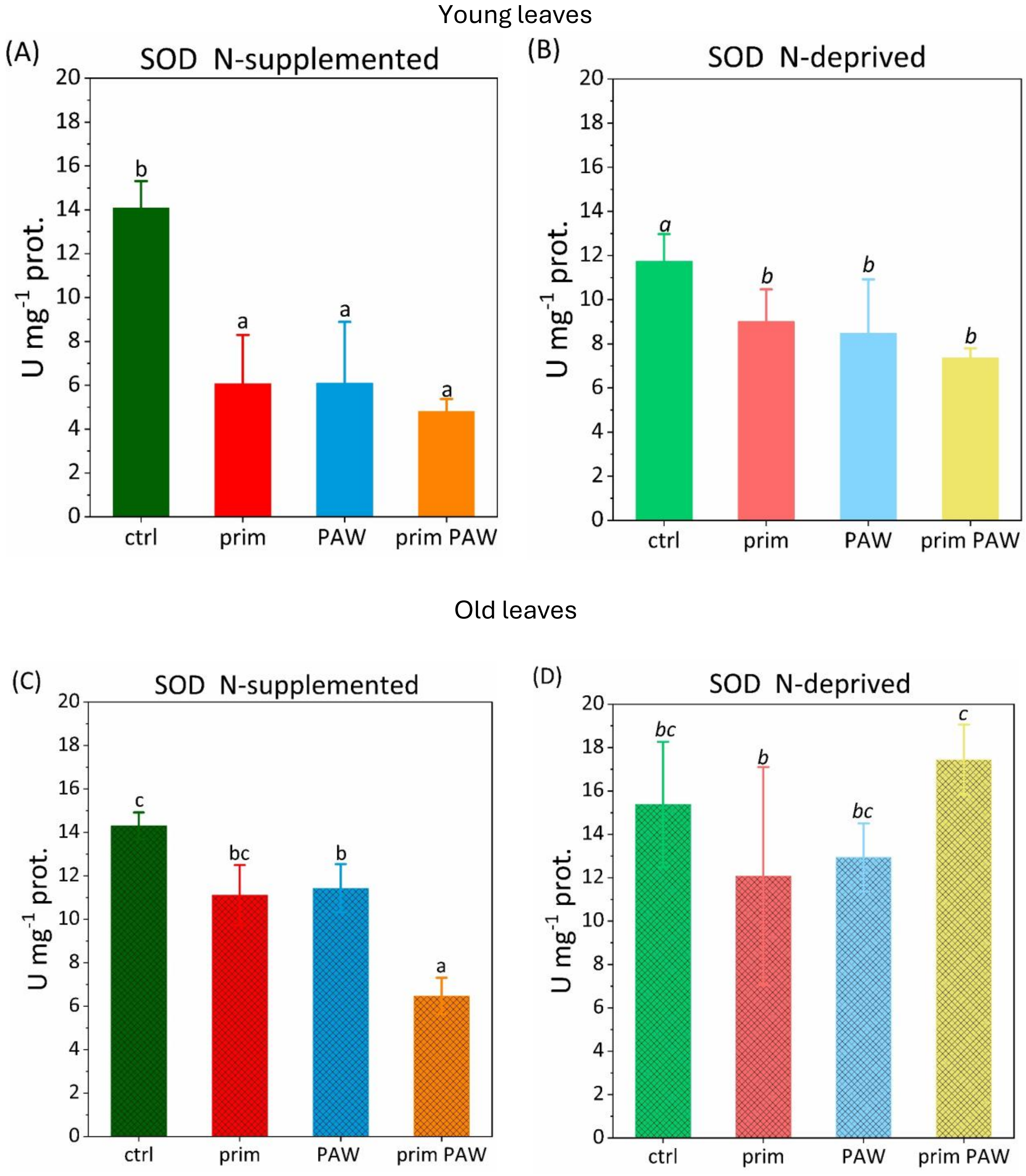

*Fig. 13 Activity of superoxide dismutase (SOD) in young (A, B) and old (C, D) leaves of lettuce cultivated in the half-strength Hoagland solution under N-supplemented (A, C) and N-deprived (B, D) conditions in combination with different PAW treatments: seed priming (prim), cultivation in PAW-activated Hoagland solution (PAW), and seed priming combined with cultivation in PAW-activated Hoagland solution (prim PAW). Values represent means ± SD. Different letters indicate significant differences at $p \leq 0.05$ (LSD test).*

There was no influence of the factor Z-treatment and N-supplementation on soluble phenolics after two-way ANOVA (Tab. 2). According to one-way ANOVA, plants cultivated with prim and PAW treatment produced more soluble phenolics than controls, independent of N supplementation. The increase under both N-supplemented and N-deprived conditions was similar, approximately 1.4-times (Fig. 14A, B).

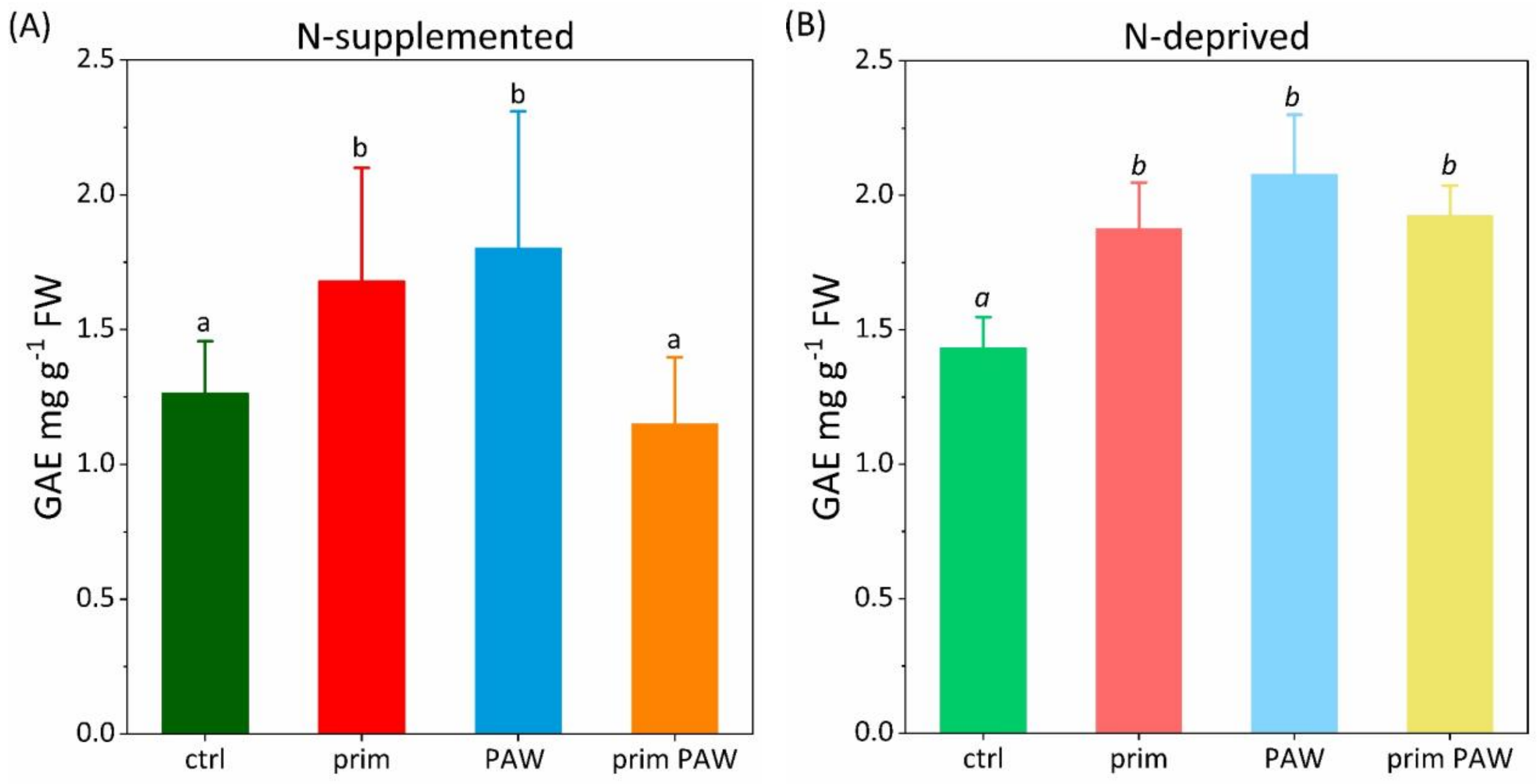

*Fig. 14 Equivalents of soluble phenolics of lettuce cultivated in the half-strength Hoagland solution under N-supplemented (A) and N-deprived (B) conditions in combination with different PAW treatments: seed priming (prim), cultivation in PAW-activated Hoagland solution (PAW), and seed priming combined with cultivation in PAW-activated Hoagland solution (prim PAW). Values represent means ± SD. Different letters indicate significant differences at $p \leq 0.05$ (LSD test).*

Nitrogen in the media and treatment influenced the content of one of the most important antioxidants, quercetin (Tab. 2). Under PAW and prim PAW cultivation, quercetin concentration increased independently of the N supplementation (Fig. 15), although under N-deprived conditions, plants produced significantly less quercetin (Fig. 15B). Under PAW treatment, plants produced 1.6-times more quercetin than control plants under N supplementation (Fig. 15A), and 1.5-times more quercetin in prim PAW in comparison with the corresponding control under N-deprivation (Fig. 15B).

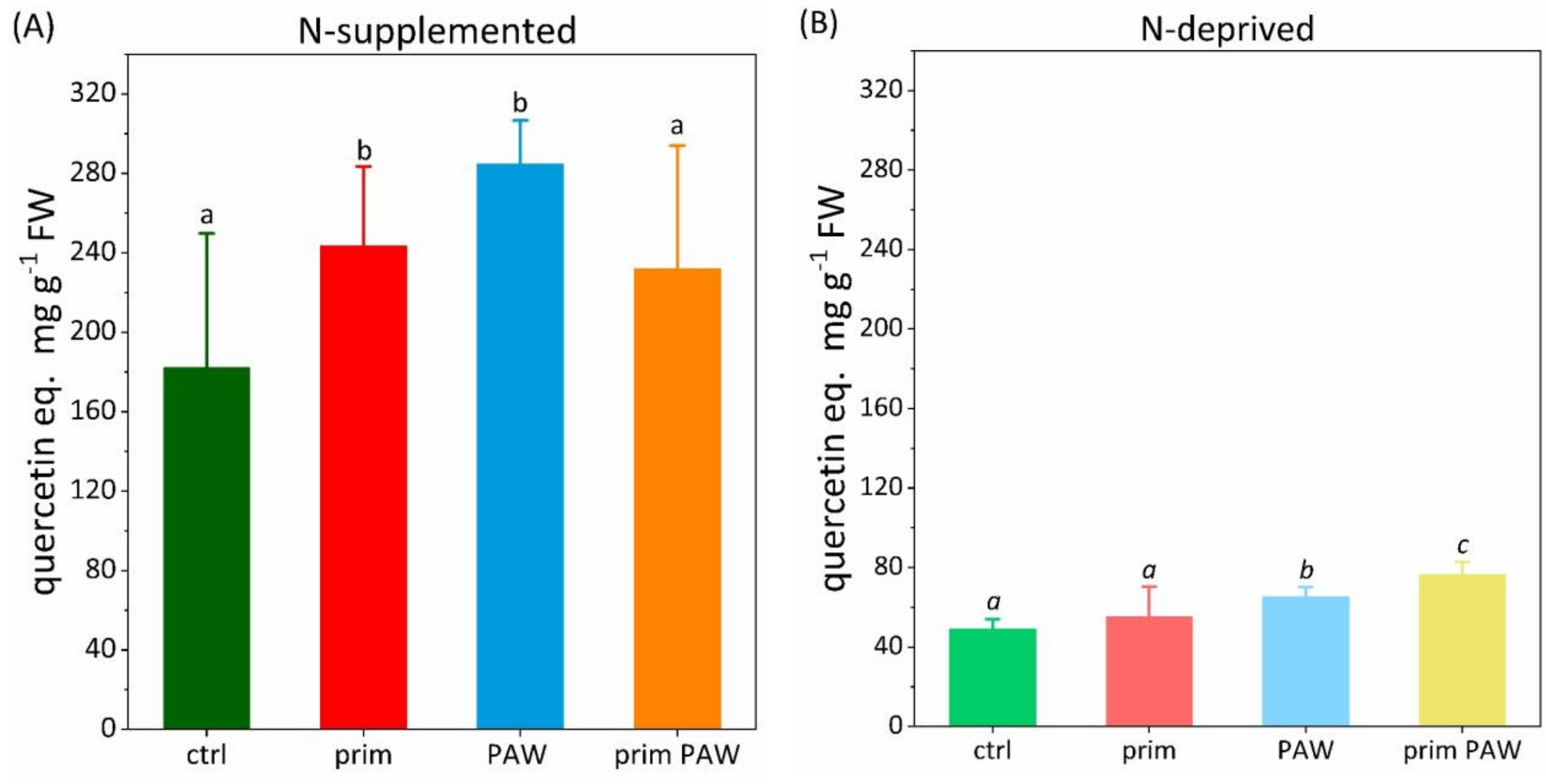

*Fig. 15 Equivalents of quercetin in lettuce cultivated in the half-strength Hoagland solution under N-supplemented (A) and N-deprived (B) conditions in combination with different PAW treatments: seed priming (prim), cultivation in PAW-activated Hoagland solution (PAW), and seed priming combined with cultivation in PAW-activated Hoagland solution (prim PAW). Values represent means ± SD. Different letters indicate significant differences at $p \leq 0.05$ (LSD test).*

**Principal component analysis**

To evaluate the combined effects of the nitrogen availability and the PAW treatment, principal component analysis (PCA) was performed separately for young and old leaves on pigments, antioxidants, and enzymes (Fig. 16).

The first two principal components (PCs) explained 69% of the total variance in young leaves (Fig. 16A). PC1 accounting for 49% of the total variability separated the evaluated samples according to the nitrogen availability in the cultivation media; all samples from N-supplemented treatments were positively correlated with PC1, whereas those from N-deprived treatments were negatively correlated with this component. The main positive loading variables contributing to PC1 were chlorophylls, pheophytins, carotenoids, and quercetin, indicating that these compounds are major drivers of variability under N-supply. SOD activity and total phenolics contributed negatively to this axis. Activity of GR, G-POX, and APX was also positively correlated with PC1; however, this correlation was less pronounced. This suggests coordinated redox regulation and pigment metabolism in N-supplemented plants.

In contrast, N-deprived samples clustered on the negative side of PC1 and were mostly associated with increased SOD activity and with total phenolics. This indicates a shift toward stress-related antioxidants under N limitation. PC2 explained the additional 20% of the total variance and mainly differentiated N-deprived PAW and prim PAW treatments, which were tightly grouped. Positive loadings on PC2 were primarily associated with GR activity, whereas APX, G-POX, and phenolics showed negative associations with this component. Overall, nitrogen availability appeared to be the dominant factor shaping the metabolic and antioxidant status of young leaves.

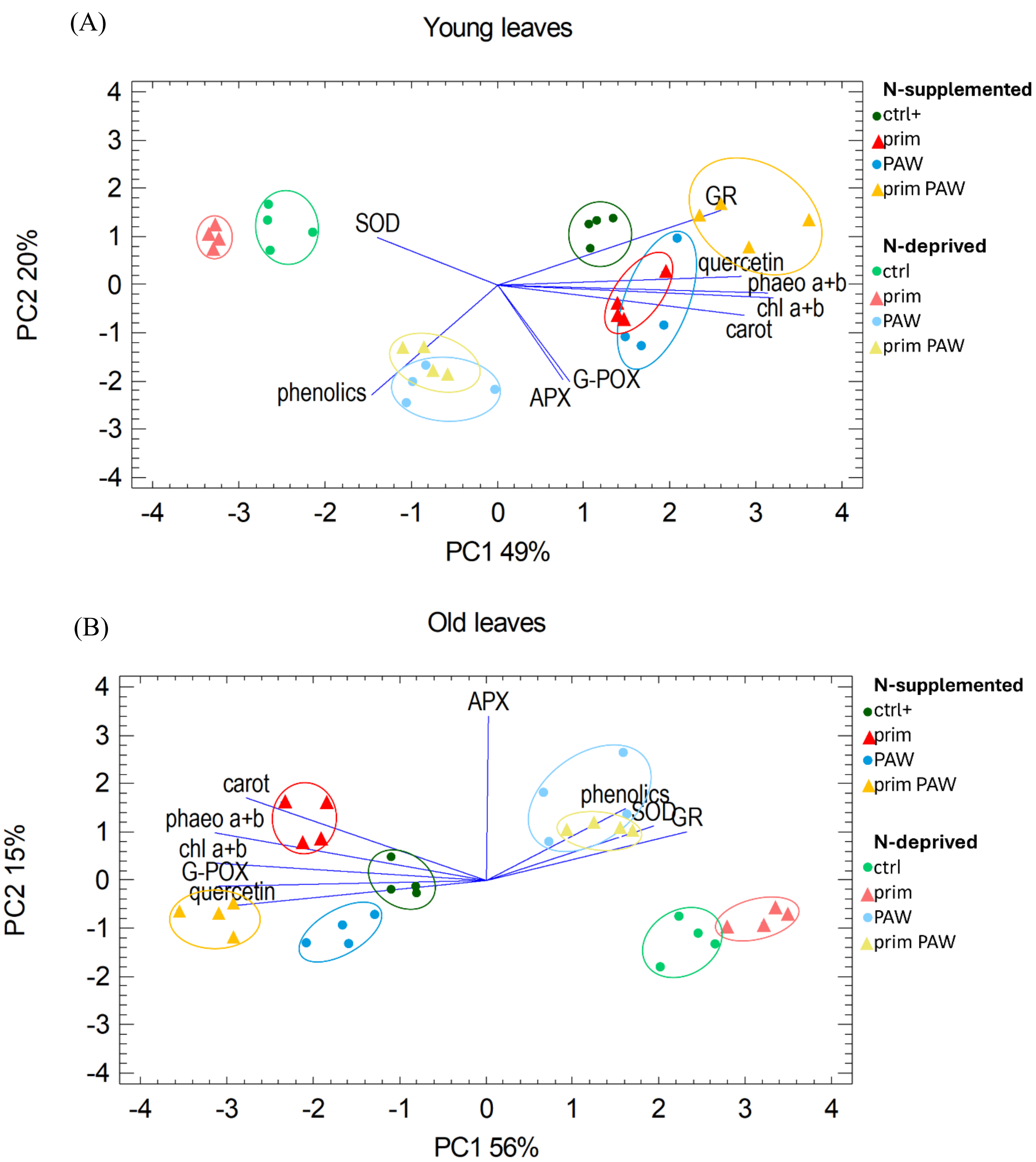

*Fig. 16 The principal component analysis (PCA) of the selected variables in young (A) and old (B) leaves of lettuce cultivated in the half-strength Hoagland solution under N-supplemented and N-deprived conditions in combination with different PAW treatments: seed priming (prim), cultivation in PAW-activated Hoagland solution (PAW), and seed priming combined with cultivation in PAW-activated Hoagland solution (prim PAW).*

The first two PCs in old leaves explained 71% of the total variability, with PC1 accounting for 56%, and PC2 for 15% (Fig. 16B). Similar to young leaves, PC1 separated N-supplemented and N-deprived plants. Positive loadings on PC1 were mainly associated with phenolics, SOD, and GR, particularly in N-deprived PAW and prim PAW-treated plants, indicating activation of antioxidant defence systems in older tissues exposed to prolonged nitrogen stress. In contrast, N-supplemented samples were negatively correlated to PC1 and associated with higher levels of chlorophylls, pheophytin, carotenoids and quercetin, reflecting better maintenance of the photosynthetic apparatus. PC2 explained an additional 15% of the total variance and was strongly associated with APX activity, which contributed predominantly to the vertical separation of samples.

The observed treatments were more distinguished in old leaves than in young leaves. PAW treatments under N-supplementation showed separation from controls along axes associated with phenolics and antioxidant enzymes, indicating that PAW may enhance redox-related adjustments in mature tissues. Under N-deprivation, prim and priming PAW treatment formed partially distinct clusters, suggesting cumulative or interaction effects of seed priming and nitrogen stress that become more pronounced with the leaf age.

**Sensory analysis of lettuce leaves**

The concentrations of reactive nitrogen species (RNS) in the samples were determined before the sensory evaluation of lettuce grown in the half-strength Hoagland solution under N-supplemented conditions (Tab. 3), with particular attention to nitrite ($NO_2^-$) levels due to their potential toxicity. Nitrite concentrations were the highest in the control and progressively decreased in the treated samples, with the lowest concentration observed in the combined priming and PAW treatment. In contrast, nitrate ($NO_3^-$) concentrations were lower in the control and increased in all treated variants, reaching the highest values in PAW-treated samples. Priming alone and in combination with PAW resulted in similarly elevated nitrate levels.

*Tab. 3 Concentration of $NO_2^-$ and $NO_3^-$ in the lettuce leaves after 10 weeks of cultivation in the half-strength Hoagland solution under N-supplemented conditions in combination with various PAW treatments: seed priming (prim), cultivation in PAW-activated Hoagland solution (PAW), and seed priming combined with cultivation in PAW activated Hoagland solution (prim PAW). Values represent means ± SD. Different letters indicate significant differences at $p \le 0.05$ (LSD test).*

| Treatment | Concentration of $NO_2^-$ (μmol $g^{-1}$ FW) | Concentration of $NO_3^-$ (μmol $g^{-1}$ FW) |
|---|---|---|
| ***ctrl*** | 0.0595 ± 0.008 c | 1.4407 ± 0.1037 a |
| ***prim*** | 0.0517 ± 0.002 b | 2.3164 ± 0.0260 b |
| ***PAW*** | 0.0488 ± 0.002 b | 2.4020 ± 0.0778 b |
| ***prim PAW*** | 0.0247 ± 0.003 a | 2.3103 ± 0.1730 b |

The sensory analysis then aimed to evaluate the effect of PAW on the subjective taste and visual quality of lettuce leaves. Control plants were rated by participants as the greenest, along with PAW-treated plants (Fig. 17). They were also perceived as the least aromatic, most bitter, and the crispiest. Lettuce cultivated with PAW was rated as the most aromatic, with the strongest vegetal flavour. Both PAW and prim PAW treatments were associated with the highest juiciness. Overall, the best performance was observed for the prim PAW treatment, which exhibited superior leaf uniformity, freshness, and sweetness, and was consistently preferred by participants as the variant they would choose in a commercial setting. The results presented in Figure 17 represent the average scores assigned by the participants, with 10 indicating the highest and 1 the lowest rating.

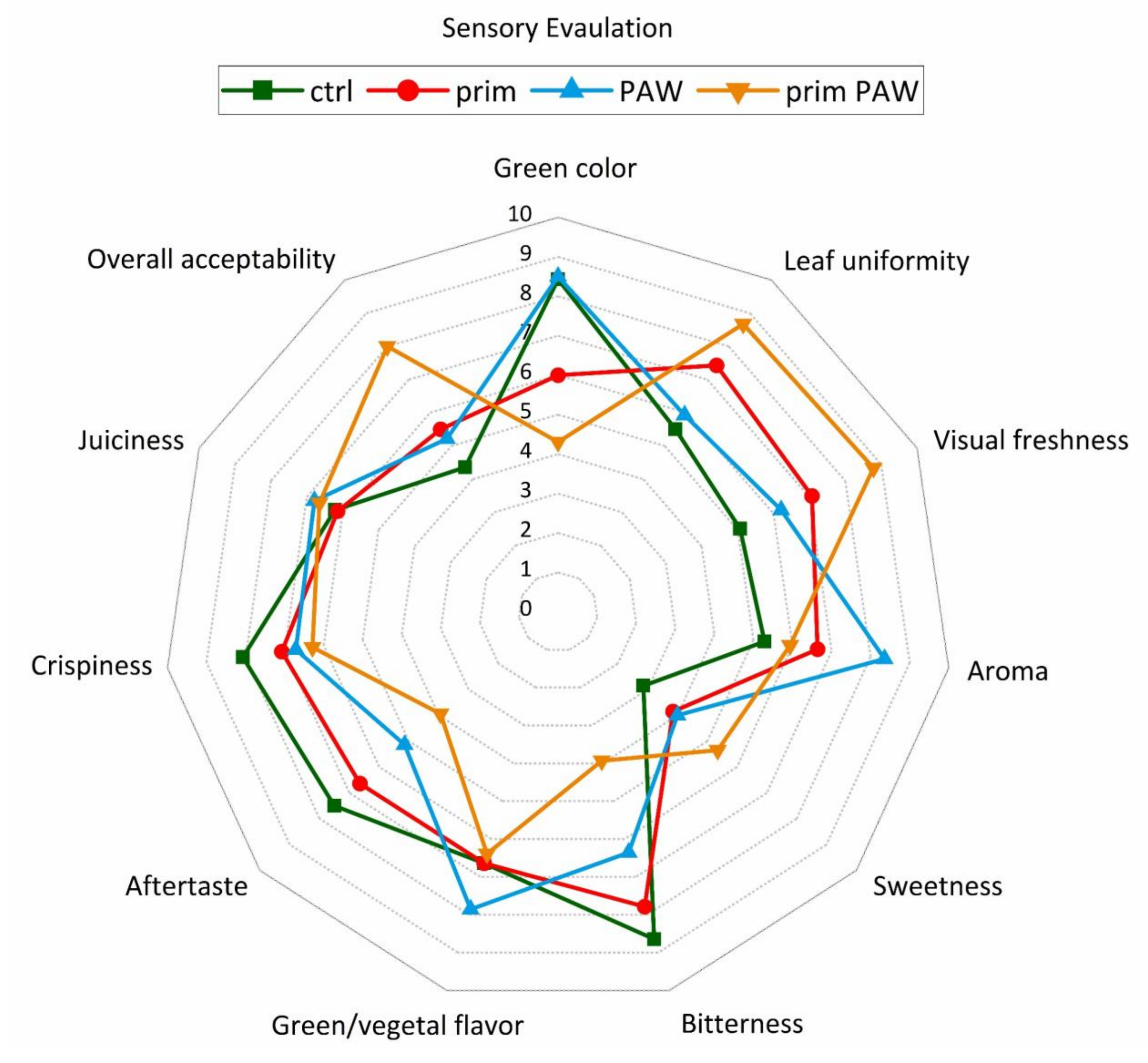

*Fig. 17 Sensory evaluation of lettuce after 10 weeks of cultivation in the half-strength Hoagland solution under N-supplemented conditions in combination with different PAW treatments: seed priming (prim), cultivation in PAW-activated Hoagland solution (PAW), and seed priming combined with cultivation in PAW-activated Hoagland solution (prim PAW).*

## Discussion

This study demonstrates that plasma-activated water (PAW) enhances the lettuce growth under hydroponic cultivation, as well as its metabolism and redox homeostasis, with effects strongly modulated by nitrogen availability. The results indicate that PAW-derived reactive oxygen and nitrogen species (RONS) play a dual role, acting both as signalling molecules and as a supplementary source of nitrogen under N-limited conditions.

Chemical characterization of PAW confirmed the presence of relatively stable nitrogen-containing species, particularly nitrate ($NO_3^-$) and nitrite ($NO_2^-$), which persisted in the solution after plasma treatment throughout the cultivation period. In contrast, hydrogen peroxide showed a decline during cultivation, reaching negligible concentrations after several days. This temporal pattern suggests that while short-lived ROS may initiate early signalling responses in plants, longer-lived RNS likely contribute to sustained effects on the plant physiology. The stability of $NO_3^-$ and $NO_2^-$ supports their potential role not only in redox signalling but also in nitrogen nutrition (Bruggeman et al., 2016). This favourable nitrogen profile, together with the presence of reactive oxygen and nitrogen species (RONS) in PAW, suggests a dual role of PAW in lettuce physiology: providing signalling molecules while also contributing to nitrogen nutrition under limited nutrient conditions (Hrgovčić et al., 2026).

In line with this, growth responses differed depending on the nitrogen availability, suggesting dual roles of PAW-derived RONS. Under N-sufficient conditions, the combined treatment of seed priming and PAW (prim PAW) resulted in the highest growth performance, whereas PAW alone did not substantially exceed control plants. In contrast, under N-deprived conditions, PAW and prim PAW treatments significantly improved the plant growth compared with untreated controls, indicating that PAW may partly compensate for nitrogen limitation. These contrasting responses suggest that when nitrogen is abundant, the beneficial effects of PAW are primarily related to signalling and priming mechanisms, whereas under nitrogen-limited conditions, nitrogen-containing species generated during plasma activation directly contribute to the plant nutrition. After uptake, nitrate is reduced through nitrate and nitrite reductases to ammonium, which is subsequently incorporated into amino acids and other nitrogen-containing metabolites essential for plant growth (Zdunek-Zastocka and Grabowska, 2019).

Reactive species present in PAW are known to influence plant hormonal signalling and redox homeostasis. Previous studies have shown that RONS may modulate abscisic acid (ABA) signalling while simultaneously stimulating gibberellin (GA)-related pathways, thereby accelerating metabolic processes associated with germination and plant growth (Yodpitak et al., 2019). Consistent with this mechanism, growth-promoting effects of PAW have been reported in hydroponically cultivated lettuce (Carmassi et al., 2022), as well as in peanut (Song et al., 2023) and radish (Rathore and Nema, 2025), largely attributed to RONS-mediated signalling pathways. Our results extend these observations by indicating that PAW may additionally serve as a supplementary nitrogen source under nutrient-limited conditions. The observed improvements in the plant growth were closely associated with changes in plant water status rather than biomass accumulation. The relatively small variation in dry weight compared to fresh weight indicates that PAW treatments primarily enhanced tissue hydration. This interpretation is further supported by sensory evaluation, in which PAW-treated plants were consistently perceived as juicier. Enhanced water retention may result from improved membrane stability and regulation of osmotic balance, processes known to be influenced by redox signalling.

Nitrogen availability strongly influenced chlorophyll levels, with nitrogen deprivation leading to a substantial decline in chlorophyll levels, reflecting the central role of nitrogen in chlorophyll biosynthesis and photosynthetic protein complexes (Agnihotri and Seth, 2016; Makino and Osmond, 1991; Stoleru et al., 2020) PAW treatments partially alleviated this decline, likely due to the presence of nitrogen-containing species that support nitrogen assimilation and pigment synthesis (Chalise et al., 2024; Ramalho et al., 2002). In addition to this nutritional effect, RONS mays contribute to the structural stability of chloroplast membranes by limiting ROS-induced oxidative damage, thereby preserving the photosynthetic efficiency (Kučerová et al., 2019; Mehrabifard et al., 2025; Ndiffo Yemeli et al., 2021; Sayed and Gadallah, 2019; Škarpa et al., 2020)

Carotenoids exhibited a different response depending on nitrogen status. Under N-supplemented conditions, PAW treatments slightly reduced carotenoid levels, whereas under N-deprivation, they significantly increased carotenoid concentrations. This suggests that PAW enhances stress-related protective mechanisms under nutrient limitation. Given the role of carotenoids in photoprotection and antioxidative defence, their accumulation likely contributes to improved resilience of the photosynthetic apparatus under stress conditions (Swapnil et al., 2021).

The antioxidant system plays a central role in plant responses to reactive oxygen and nitrogen species generated during plasma activation. In the present study, PAW treatments modulated several antioxidant enzymes, particularly those involved in the ascorbate–glutathione cycle, suggesting activation of ROS detoxification pathways. Enhanced antioxidant activity is widely recognized as an important factor contributing to the plant growth and vitality (Hasanuzzaman et al., 2012), and strengthening of the antioxidative system represents an important adaptive mechanism in plants (Fatma et al., 2021).

The higher activity of ascorbate peroxidase (APX) observed particularly in young leaves of PAW-treated plants indicates activation of hydrogen peroxide detoxification pathways. APX is a key enzyme in the ascorbate–glutathione cycle and is rapidly induced in response to ROS signalling, helping to maintain cellular redox balance. Similar responses have been reported in tomato seedlings, where low external concentrations of hydrogen peroxide stimulated APX activity and improved stress tolerance (İşeri et al., 2013). Glutathione reductase showed contrasting responses between young and old leaves, reflecting differences in redox metabolism between developing and mature tissues. Because GR maintains reduced glutathione pools required for antioxidant defense, changes in its activity may indicate adjustments in cellular redox homeostasis under PAW treatments. In contrast, SOD activity remained largely unchanged, suggesting that superoxide production did not exceed the basal detoxification capacity and that downstream antioxidant enzymes played a more prominent role in maintaining redox balance. Similar patterns have been reported in other crop species including barley, maize, and wheat (Kučerová et al., 2019; Ndiffo Yemeli et al., 2021).

Changes in secondary metabolites further support the role of PAW in modulating plant metabolism. Although total soluble phenolics showed only moderate increases, their accumulation in PAW-treated plants suggests activation of defence-related metabolic pathways. This response is consistent with the dual role of ROS, which can stimulate phenolic biosynthesis (Fan et al., 2020; Iranbakhsh et al., 2018, 2017; Okruhlicová et al., 2025; Tappi et al., 2018) and promote their oxidation (Ramazzina et al., 2015), resulting in a balanced net effect.

In contrast, the flavonoid quercetin showed a more pronounced response to PAW treatments with increased concentrations observed regardless of nitrogen availability. Quercetin is a potent antioxidant capable of scavenging free radicals and contributing to cellular redox homeostasis, while also influencing the characteristic bitterness of lettuce leaves (Anand David AV et al., 2016; Pandey et al., 2022). Increased quercetin concentrations in PAW and prim PAW treatments indicate that plasma-derived reactive species may stimulate the biosynthesis of specific flavonoids involved in antioxidant defence. Conversely, the reduced quercetin levels observed under N-deprived conditions are consistent with previous reports indicating that nitrogen limitation can restrict the synthesis of certain secondary metabolites. Additionally, reactive species present in PAW may facilitate the release of bound phenolic compounds from cell wall structures, further contributing to the observed increase in polyphenol levels (Adom and Liu, 2002).

Principal component analysis (PCA) confirmed that nitrogen availability is the primary factor creating metabolic and antioxidative profiles. In young leaves, nitrogen status clearly separated samples, with N-supplemented plants associated with higher pigment content and metabolic activity, while N-deprived plants showed stronger association with stress-related antioxidants. In old leaves, the separation was more complex, reflecting cumulative metabolic adjustments and stronger treatment effects. PAW-treated plants, particularly under N-deprivation, showed distinct clustering patterns, indicating coordinated modulation of antioxidant systems and secondary metabolism.

Since sensory analysis of the lettuce leaves including taste parameters is a part of this study, in addition to PAW composition, nitrate ($NO_3^-$) and nitrite ($NO_2^-$) concentrations were also determined in the edible leaf tissue. Nitrate levels ranged from 1.4 to 2.4 μmol $g^{-1}$ fresh weight (approximately 87–149 mg $kg^{-1}$ fresh weight), while nitrite concentrations were considerably lower, ranging from 0.02 to 0.06 μmol $g^{-1}$ fresh weight (approximately 0.9–2.8 mg $kg^{-1}$). These values are well below the maximum limits established by European Union regulations for leafy vegetables (2000–5000 nitrite mg $kg^{-1}$ fresh weight, depending on season and cultivation conditions) and are comparable to concentrations commonly found in commercially available products (Mortensen et al., 2017b, 2017a).

According to the European Food Safety Authority (EFSA), the acceptable daily intake (ADI) is established at 3.7 mg $kg^{-1}$ body weight per day for nitrate and 0.07 mg $kg^{-1}$ body weight per day for nitrite, based on evaluations by the Scientific Committee on Food and JECFA (Mortensen et al., 2017b, 2017a). Based on a model consumption of 100 g of lettuce per day, the estimated intake would correspond to approximately 8.7–14.9 mg of nitrate and 0.09–0.28 mg of nitrite, representing only a small fraction of the ADI values for an average adult. These results indicate that PAW treatment does not pose a risk in terms of nitrate or nitrite accumulation in edible plant tissues, supporting its safe application in hydroponic lettuce cultivation.

Overall, the results of this study demonstrate that PAW influences lettuce plant physiology through multiple interconnected mechanisms involving redox signalling, nitrogen metabolism, and stress response pathways. The interplay between these processes determines the final phenotypic outcome, which varies depending on nutrient availability and tissue developmental stage.

## Conclusion

*Lactuca sativa* plants were grown hydroponically under the N-sufficient and N-deprived half-strength Hoagland solutions. We evaluated four treatments with plasma-activated water: control, seed priming in the PAW-activated Hoagland, PAW-activated Hoagland cultivation, and their combination under two nutrient regimes.

The results demonstrate that nitrogen availability is the dominant factor shaping metabolic and antioxidative profiles in both young and old leaves in lettuce, while PAW treatments act as modulators of these processes.

Principal component analysis revealed a clear nitrogen-dependent separation of samples, confirming that nutrient status defines the primary metabolic framework. However, the plant tissue age significantly influenced the nature of the response. Young leaves were characterized by a stronger association of N-deprivation with SOD activity and phenolic accumulation, whereas old leaves exhibited a broader coordination of antioxidant enzymes, including SOD and GR, indicating progressive redox remodelling during the leaf maturation.

PAW treatments caused more distinct and structured effects in old leaves than in young leaves, suggesting that the impact of plasma-derived reactive species becomes more pronounced over time as metabolic adjustments accumulate. These effects were particularly evident under N-deprived conditions, where PAW partially compensated for nutrient limitation by enhancing growth, pigment content, and antioxidant capacity.

Taken together, the findings support the hypothesis that PAW-derived reactive species have a dual function: they act as signalling molecules regulating redox homeostasis and metabolic activity and simultaneously serve as an additional source of nitrogen through stable compounds, such as nitrate and nitrite. This dual role enables PAW to improve plant performance under optimal conditions and mitigate the negative effects of N-deprivation.

Our results highlight the potential of PAW as a sustainable tool in crop production, particularly in systems with reduced fertilizer input, where it may enhance both plant growth and stress resilience.

**Acknowledgment**

This work was funded by the EU NextGenerationEU through the Recovery and Resilience Plan for Slovakia under the project No. 09I03-03-V03-00033 EnvAdwice, and Slovak Research and Development Agency APVV-22-0247.